\documentclass[aps,prd,preprint,floats,epsf,superscriptaddress,nofootinbib]{revtex4-1}

\usepackage{graphicx} 
\usepackage{amsmath,amssymb,amsfonts,mathtools} 
\usepackage{mathrsfs}
\usepackage{slashed} 
\usepackage{braket} 
\usepackage{indentfirst}
\usepackage{xcolor}
\usepackage{dsfont}
\usepackage[normalem]{ulem}
\usepackage{dcolumn}
\usepackage{bm}
\usepackage{bbm}
\usepackage{multirow}
\usepackage{colortbl}
\usepackage{adjustbox}
\usepackage{float}
\usepackage[T1]{fontenc}
\usepackage{calligra}
\usepackage[T1]{pbsi}
\usepackage{CJKutf8} 

\usepackage[colorlinks,
            linkcolor=black,
            anchorcolor=black,
            citecolor=black
            ]{hyperref} 

\newcommand{\comment}[1]{} 

\def\bwt{\begin{widetext}}
\def\ewt{\end{widetext}}
\def\be{\begin{equation}}
\def\ee{\end{equation}}
\def\bea{\begin{eqnarray}}
\def\eea{\end{eqnarray}}
\def\bean{\begin{eqnarray*}}
\def\eean{\end{eqnarray*}}
\def\bary{\begin{array}}
\def\eary{\end{array}}

\def\bit{\begin{itemize}}
\def\eit{\end{itemize}}

\newcommand{\cellg}{\cellcolor{green!30}}
\newcommand{\celly}{\cellcolor{yellow!30}}
\newcommand{\cellr}{\cellcolor{red!30}}

\begin{document}


\bigskip

\title{Updated constraints on Georgi-Machacek model, and its electroweak phase transition and associated gravitational waves}

\author{Ting-Kuo Chen}
\email[e-mail: ]{tkchen@phys.ntu.edu.tw}
\affiliation{Department of Physics, National Taiwan University, Taipei, Taiwan 10617, R.O.C.}

\author{Cheng-Wei Chiang}
\email[e-mail: ]{chengwei@phys.ntu.edu.tw}
\affiliation{Department of Physics, National Taiwan University, Taipei, Taiwan 10617, R.O.C.}
\affiliation{Physics Division, National Center for Theoretical Sciences, Taipei, Taiwan 10617, R.O.C.}

\author{Cheng-Tse Huang}
\email[e-mail: ]{r09222065@ntu.edu.tw}
\affiliation{Department of Physics, National Taiwan University, Taipei, Taiwan 10617, R.O.C.}

\author{Bo-Qiang Lu}
\email[e-mail: ]{bqlu@zjhu.edu.cn}
\affiliation{School of Science, Huzhou University, Huzhou, Zhejiang 313000, P.R.O.C.}

\date{\today}

\begin{abstract}
With theoretical constraints such as perturbative unitarity and vacuum stability conditions and updated experimental data of Higgs measurements and direct searches for exotic scalars at the LHC, we perform an updated scan of the allowed parameter space of the Georgi-Machacek (GM) model.  With the refined global fit, we examine the allowed parameter space for inducing strong first-order electroweak phase transitions (EWPTs) and find only the one-step phase transition is phenomenologically viable.  Based upon the result, we study the associated gravitational wave (GW) signals and find most of which can be detected by several proposed experiments.  We also make predictions on processes that may serve as promising probes to the GM model in the near future at the LHC, including the di-Higgs productions and several exotic scalar production channels.
\end{abstract}


\maketitle


\setlength\parindent{15pt} 
\setlength{\parskip}{10pt} 

\newpage

\section{Introduction}
\label{sec:Introduction}
The discovery of the 125-GeV scalar resonance at the LHC~\cite{ATLAS:2012yve,CMS:2012qbp} has claimed its consistency with the Standard Model (SM) Higgs boson in terms of particle content. Nonetheless, as there remain several experimental observations that ask for new physics explanations, the exact structure of the electroweak sector is still under intense exploration. one example is the deviations from the SM predictions for the ${hff}$, ${hVV}$, ${hZ\gamma}$ and ${h\gamma\gamma}$ couplings, as given in Refs.~\cite{CMS:2018uag,ATLAS:2019nkf}, that still allow a beyond-SM interpretation. Another example is the electroweak baryogenesis problem, the success of which requires the occurrence of a strong first-order electroweak phase transition (EWPT). However, according to the non-perturbative lattice computations~\cite{Kajantie:1996mn,Gurtler:1997hr,Csikor:1998eu}, the electroweak symmetry breaking (EWSB) of the SM only occurs through a smooth crossover transition around the temperature $T\sim100$~GeV.  Thus, extensions to the SM Higgs sector are called for.

In this work, we study the Georgi-Machacek (GM) model~\cite{Georgi:1985nv,Chanowitz:1985ug}, which introduces one complex and one real scalar triplets that preserves the custodial symmetry at tree level after the electroweak symmetry breakdown (EWSB). The model predicts the existence of several Higgs multiplets, whose mass eigenstates form one quintet ($H_5$), one triplet ($H_3$), and two singlets ($H_1$ and $h$) under the custodial symmetry, thus leading to rich Higgs phenomenology. For example, enhancements in the $hWW$ and $hZZ$ couplings compared to the SM predictions can be achieved through the additional triplet-gauge interactions, and considerable deviations from the SM predictions for the di-Higgs production rates can also be induced through the modification to the Higgs self-couplings as well as the new contribution from $H_1$ through the singlet mixing.  The model also has the capability of providing Majorana mass to neutrinos through the triplet vacuum expectation values (VEVs).  Moreover, as we show in this study, the GM model can generate strong first-order EWPTs while satisfying all the current collider measurement constraints in certain phase space, and can further lead to detectable stochastic gravitational wave (GW) backgrounds through the bubble dynamics between the symmetric and broken phases~\cite{Caprini:2015zlo,Caprini:2019egz}. These salient features of the model arouse in recent years a series of studies on collider phenomenology~\cite{Chiang:2013rua,Chiang:2012cn,Chiang:2014bia,Chiang:2015kka,Chiang:2015amq,Chiang:2015rva,Logan:2015xpa,Degrande:2017naf,Logan:2017jpr,Chang:2017niy} as well as the EWPT~\cite{Chiang:2014hia,Zhou:2018zli}.

To explore the phase space of the GM model that satisfies essential theoretical bounds and experimental constraints from various LHC and Tevatron measurements, we perform Bayesian Markov-Chain Monte Carlo (MCMC) global fits in the model with \texttt{HEPfit}~\cite{DeBlas:2019ehy}. Compared with the previous work~\cite{Chiang:2018cgb}, we have updated the experimental data and refined several fitting setups to achieve more restraining results. With the parameter samples extracted from the phase space that satisfies all the mentioned constraints, we go on to calculate the EWPT characteristics by employing a high-temperature approximation for the thermal effective potential, and predict the GW backgrounds induced from the bubble dynamics.

The structure of this paper is as follows.  In Sec.~\ref{sec:The Georgi-Machacek Model}, we review the GM model and give the theoretical constraints to be imposed on the model.  In Sec.~\ref{sec:Constraints}, we choose the model Lagrangian parameters as our scanning parameters and set their prior distributions.  We then show step by step how various theoretical and experimental constraints restrict the parameter space.  Based on the scanning result, we further find the parameter sets that will lead to sufficiently strong first-order EWPTs in Sec.~\ref{sec:Electroweak Phase Transition and Gravitational Waves}.  We calculate the associated GW spectra and make a comparison with the sensitivities of several proposed GW experiments.  Moreover, we use these parameter sets to make predictions for the most promising constraining/discovering modes at the LHC in Sec.~\ref{sec:Predictions}. Finally, we discuss and summarize our findings in Sec.~\ref{sec:Discussions and Summary}.

\section{The Georgi-Machacek Model}
\label{sec:The Georgi-Machacek Model}

The electroweak (EW) sector of the GM model comprises one isospin doublet scalar field with hypercharge $Y=1 / 2$\footnote{We adopt the hypercharge convention such that $Q=T_3+Y$.}, one complex isospin triplet scalar field with $Y=1,$ and one real isospin triplet scalar field with $Y=0$. These fields are denoted respectively by\footnote{The sign conventions for the charge conjugate fields are $\phi^-=(\phi^+)^*$, $\chi^-=(\chi^+)^*$, $\xi^-=(\xi^+)^*$ and $\chi^{--}=(\chi^{++})^*$.}
\begin{equation}\label{eq:scalar field}
\begin{array}{l}
\phi=\left(\begin{array}{c}
\phi^{+} \\
\phi^{0}
\end{array}\right), \quad \chi=\left(\begin{array}{c}
\chi^{++} \\
\chi^{+} \\
\chi^{0}
\end{array}\right), \quad \xi=\left(\begin{array}{c}
\xi^{+} \\
\xi^{0} \\
-\left(\xi^{+}\right)^{*}
\end{array}\right)
\end{array} ~,
\end{equation}
where the neutral components before the EWSB are parametrized as $\phi^{0}=\left(h_{\phi}+i a_{\phi}\right)/\sqrt{2}$, $\chi^{0}=\left(h_{\chi}+i a_{\chi}\right)/\sqrt{2}$, and $\xi^{0}=h_{\xi}$. A global $\mathrm{SU}(2)_{L} \times \mathrm{SU}(2)_{R}$ symmetry, which is explicitly broken by the Yukawa and the hypercharge-$U(1)$ gauge interactions, is imposed on the Higgs potential at tree level, which can be succinctly expressed by introducing the $\mathrm{SU}(2)_{L} \times \mathrm{SU}(2)_{R}$-covariant forms of the fields:
\begin{equation}
\begin{aligned}
\Phi & \equiv\left(\epsilon_{2} \phi^{*}, \phi\right)=\left(\begin{array}{ccc}
\left(\phi^{0}\right)^{*} & \phi^{+} \\
-\left(\phi^{+}\right)^{*} & \phi^{0}
\end{array}\right), \quad \text {with } \epsilon_{2}=\left(\begin{array}{cc}
0 & 1 \\
-1 & 0
\end{array}\right) ~, \\
\Delta & \equiv\left(\epsilon_{3} \chi^{*}, \xi, \chi\right)=\left(\begin{array}{ccc}
\left(\chi^{0}\right)^{*} & \xi^{+} & \chi^{++} \\
-\left(\chi^{+}\right)^{*} & \xi^{0} & \chi^{+} \\
\left(\chi^{++}\right)^{*} & -\left(\xi^{+}\right)^{*} & \chi^{0}
\end{array}\right), \quad \text { with } \epsilon_{3}=\left(\begin{array}{ccc}
0 & 0 & 1 \\
0 & -1 & 0 \\
1 & 0 & 0
\end{array}\right) ~.
\end{aligned}
\end{equation}

The Lagrangian of the EW sector is given by
\begin{equation}
\mathcal{L}=\frac{1}{2} \operatorname{tr}\left[\left(D^{\mu} \Phi\right)^{\dagger} \left(D_{\mu} \Phi\right)\right]+\frac{1}{2} \operatorname{tr}\left[\left(D^{\mu} \Delta\right)^{\dagger} \left(D_{\mu} \Delta\right)\right]-V(\Phi, \Delta) ~,
\end{equation}
with the most general potential invariant under the gauge and global $\mathrm{SU}(2)_{L} \times \mathrm{SU}(2)_{R} \times \mathrm{U(1)_Y}$ symmetries as
\begin{equation}
\begin{aligned}
V(\Phi, \Delta)=& \frac{1}{2} m_{1}^{2} \operatorname{tr}\left[\Phi^{\dagger} \Phi\right]+\frac{1}{2} m_{2}^{2} \operatorname{tr}\left[\Delta^{\dagger} \Delta\right]+\lambda_{1}\left(\operatorname{tr}\left[\Phi^{\dagger} \Phi\right]\right)^{2}+\lambda_{2}\left(\operatorname{tr}\left[\Delta^{\dagger} \Delta\right]\right)^{2} \\
&+\lambda_{3} \operatorname{tr}\left[\left(\Delta^{\dagger} \Delta\right)^{2}\right]+\lambda_{4} \operatorname{tr}\left[\Phi^{\dagger} \Phi\right] \operatorname{tr}\left[\Delta^{\dagger} \Delta\right]+\lambda_{5} \operatorname{tr}\left[\Phi^{\dagger} \frac{\sigma^{a}}{2} \Phi \frac{\sigma^{b}}{2}\right] \operatorname{tr}\left[\Delta^{\dagger} T^{a} \Delta T^{b}\right] \\
&+\mu_{1} \operatorname{tr}\left[\Phi^{\dagger} \frac{\sigma^{a}}{2} \Phi \frac{\sigma^{b}}{2}\right]\left(P^{\dagger} \Delta P\right)_{a b}+\mu_{2} \operatorname{tr}\left[\Delta^{\dagger} T^{a} \Delta T^{b}\right]\left(P^{\dagger} \Delta P\right)_{a b} ~,
\end{aligned}
\end{equation}
where $\sigma^a$ and $T^a$ are the $2\times2$ and $3\times3$ representations of the $\mathrm{SU}(2)$ generators, and the matrix $P$, which rotates $\Delta$ into the Cartesian basis, is given by
\begin{equation*}
P=\frac{1}{\sqrt{2}}\left(\begin{array}{ccc}
-1 & i & 0 \\
0 & 0 & \sqrt{2} \\
1 & i & 0
\end{array}\right) ~.
\end{equation*}

The vacuum potential is given by
\begin{equation}\label{potential at 0}
V_0=\frac{m_{1}^{2}}{2} v_{\phi}^{2}+\frac{3}{2}m_{2}^{2} v_{\Delta}^{2}+\lambda_{1} v_{\phi}^{4}+\frac{3}{2}\left(2 \lambda_{4}+\lambda_{5}\right) v_{\phi}^{2} v_{\Delta}^{2}+3\left(\lambda_{3}+3 \lambda_{2}\right) v_{\Delta}^{4}+\frac{3}{4} \mu_{1} v_{\phi}^{2} v_{\Delta}+6 \mu_{2} v_{\Delta}^{3} ~,
\end{equation} \label{eq:vacuum alignment}
where the VEVs\footnote{As elucidated in Ref.~\cite{Chen:2022ocr}, one has to choose ``aligned'' triplet VEVs for the custodially symmetric potential.  Assuming misaligned VEVs would lead to undesirable Goldstone and tachyonic modes in the model.}
\begin{equation}\label{eq:custudial symmetry}
\left\langle h_{\phi}\right\rangle=v_{\Phi}, \quad \left\langle h_{\chi}\right\rangle= \sqrt{2} v_{\Delta}, \quad \left\langle h_{\xi}\right\rangle=v_{\Delta} 
\end{equation}
preserve the custodial $\mathrm{SU}(2)_V$ symmetry by breaking the $\mathrm{SU}(2)_L\times \mathrm{SU}(2)_R$ symmetry diagonally, and satisfy $v=\sqrt{v_\phi^2+8v_\Delta^2}\simeq 246$~GeV. The tadpole conditions are given by
\begin{equation}
\frac{\partial V(\Phi, \Delta)}{\partial h_{\phi}}\Bigg\vert_0=\frac{\partial V(\Phi, \Delta)}{\partial h_{\chi}}\Bigg\vert_0=\frac{\partial V(\Phi, \Delta)}{\partial h_{\xi}}\Bigg\vert_0=0 ~.
\end{equation}
Since the last two conditions are equivalent, we eventually have two linearly independent conditions:
\begin{equation}\label{eq:tadpole conditions}
\begin{aligned}
m_{1}^{2} &=-4 \lambda_{1} v_{\Phi}^{2}-6 \lambda_{4} v_{\Delta}^{2}-3 \lambda_{5} v_{\Delta}^{2}-\frac{3}{2} \mu_{1} v_{\Delta} ~, \\
m_{2}^{2} &=-12 \lambda_{2} v_{\Delta}^{2}-4 \lambda_{3} v_{\Delta}^{2}-2 \lambda_{4} v_{\Phi}^{2}-\lambda_{5} v_{\Phi}^{2}-\mu_{1} \frac{v_{\Phi}^{2}}{4 v_{\Delta}}-6 \mu_{2} v_{\Delta} ~.
\end{aligned}
\end{equation}
We further define
\begin{equation}
M_{1}^{2} \equiv-\frac{v}{\sqrt{2} \cos \beta} \mu_{1}, \quad M_{2}^{2} \equiv-3 \sqrt{2} \cos \beta v \mu_{2}
\end{equation}
to simplify the notations, where $\tan{\beta}=v_\phi/\left(2\sqrt{2}v_\Delta\right)$.

Before we discuss the mass spectrum of the scalars, it is convenient to classify them according to their custodial $\mathrm{SU}(2)_V$ isospins. We decompose the $\mathbf{2}\otimes\mathbf{2}$ representation $\Phi$ and the $\mathbf{3}\otimes\mathbf{3}$ representation $\Delta$ into irreducible $\mathbf{1}\oplus\mathbf{3}$ and $\mathbf{1}\oplus\mathbf{3}\oplus\mathbf{5}$ representations, respectively. In general, the two singlet fields and the two triplet fields can further mix respectively with each other, and three Nambu-Goldstone (NG) modes to be eaten by the weak gauge bosons are produced from the latter mixing. The physical quintet $(H_5^{\pm\pm},H_5^{\pm},H_5^{0})$, the physical triplet $(H_3^{\pm}, H_3^{0})$, and the two physical singlets $(H_1,h)$ can be related to the original fields via
\begin{equation}\label{flavor eigenstates}
\begin{array}{l}
H_{5}^{++}=\chi^{++}, \quad H_{5}^{+}=\frac{1}{\sqrt{2}}\left(\chi^{+}-\xi^{+}\right), \quad H_{5}^{0}=\sqrt{\frac{1}{3}} h_{\chi}-\sqrt{\frac{2}{3}} h_{\xi}, \\
H_{3}^{+}=-\cos \beta \phi^{+}+\sin \beta \frac{1}{\sqrt{2}}\left(\chi^{+}+\xi^{+}\right), \quad H_{3}^{0}=-\cos \beta a_{\phi}+\sin \beta a_{\chi}, \\
h=\cos \alpha h_{\phi}-\frac{\sin \alpha}{\sqrt{3}}\left(\sqrt{2} h_{\chi}+h_{\xi}\right), \quad H_{1}=\sin \alpha h_{\phi}+\frac{\cos \alpha}{\sqrt{3}}\left(\sqrt{2} h_{\chi}+h_{\xi}\right)~,
\end{array}
\end{equation}
where the mixing angle $\alpha \in \left( -\pi/2 , \pi/2 \right)$ is given by
\begin{equation}\label{alpha}
\tan 2 \alpha=\frac{2\left(M^{2}\right)_{12}}{\left(M^{2}\right)_{22}-\left(M^{2}\right)_{11}} ~,
\end{equation}
with
\begin{equation}
\begin{array}{l}
\left(M^{2}\right)_{11}= 8 \lambda_1 v_\phi^2 =8 \lambda_{1} v^{2} \sin ^{2} \beta ~, \\
\left(M^{2}\right)_{22}=\left(3 \lambda_{2}+\lambda_{3}\right) v^{2} \cos ^{2} \beta+M_{1}^{2} \sin ^{2} \beta-\frac{1}{2} M_{2}^{2} ~, \\
\left(M^{2}\right)_{12}=\sqrt{\frac{3}{2}} \sin \beta \cos \beta\left[\left(2 \lambda_{4}+\lambda_{5}\right) v^{2}-M_{1}^{2}\right] ~.
\end{array}
\end{equation}
The mass eigenvalues are then given by
\begin{equation}\label{masses}
\begin{aligned}
m_{H_{5}}^{2} & \equiv m_{H_{5}^{\pm\pm}}^{2}=m_{H_{5}^{\pm}}^{2}=m_{H_{5}^{0}}^{2}=\left(M_{1}^{2}-\frac{3}{2} \lambda_{5} v^{2}\right) \sin ^{2} \beta+\lambda_{3} v^{2} \cos ^{2} \beta+M_{2}^{2} ~, \\
m_{H_{3}}^{2} & \equiv m_{H_{3}^{\pm}}^{2}=m_{H_{3}^{0}}^{2}=M_{1}^{2}-\frac{1}{2} \lambda_{5} v^{2} ~, \\
m_{H_{1}}^{2} & =\left( M^{2}\right)_{11} \sin ^{2} \alpha+\left(M^{2}\right)_{22} \cos ^{2} \alpha+2 \left(M^{2}\right)_{12} \sin \alpha \cos \alpha ~, \\
m_{h}^{2} & =\left(M^{2}\right)_{11} \cos ^{2} \alpha+\left(M^{2}\right)_{22} \sin ^{2} \alpha-2 \left(M^{2}\right)_{12} \sin \alpha \cos \alpha ~,
\end{aligned}
\end{equation}
where we identify $h$ as the 125-GeV SM-like Higgs. We remark here that because of the preserved custodial symmetry at tree level, the quintet and triplet mass spectra are degenerate, respectively.

The first thing we now observe is the modification to the trilinear Higgs self-coupling, which is given by
\begin{equation}
\begin{aligned}
g_{h h h} =&
24 \cos^{3}\alpha \lambda_{1} v_{\phi}+6 \cos \alpha \sin^{2}\alpha v_{\phi}\left(2 \lambda_{4}+\lambda_{5}\right)\\&+\frac{3}{2} \sqrt{3} \cos^{2}\alpha \sin\alpha\left[4 v_{\Delta}\left(-2 \lambda_{4}-\lambda_{5}\right)-\mu_{1}\right] -4 \sqrt{3} \sin^{3}\alpha\left[\mu_{2}+2 v_{\Delta}\left(3 \lambda_{2}+\lambda_{3}\right)\right] ~,
\end{aligned}
\end{equation}
where the SM counterpart is given by $g^{\rm SM}_{hhh}=3m_h^2/v$. On the other hand, the singlet mixing also leads to
\begin{equation}
\begin{aligned}
g_{H_{1} h h}=& 24 \lambda_{1} \cos^2{\alpha} \sin{\alpha} v_{\phi}+8 \sqrt{3} \cos{\alpha} \sin^2{\alpha} v_{\Delta}\left(\lambda_{3}+3 \lambda_{2}\right)\\
&+2\left[\sqrt{3} \cos{\alpha} v_{\Delta}\left(3 \cos^2{\alpha}-2\right)+\sin{\alpha} v_{\phi}\left(1-3 \cos^2{\alpha}\right)\right]\left(2 \lambda_{4}+\lambda_{5}\right) \\
&+\frac{\sqrt{3}}{2} \mu_{1} \cos{\alpha}\left(3 \cos^2{\alpha}-2\right)+4 \sqrt{3} \mu_{2} \cos{\alpha} \sin^2{\alpha} ~.
\end{aligned}
\end{equation}
Because of these two couplings, the di-Higgs production rate predicted by the GM model can be considerably different from the SM prediction, making it one of the most interesting channels to be studied.

Moreover, the couplings of $h$ to the SM fermions $f$ and weak gauge bosons $V = W,Z$ are modified respectively as
\begin{equation}
\begin{aligned}
g_{h f \bar{f}} & =\kappa_F \times g_{h f \bar{f}}^{\mathrm{SM}}~,\\ 
g_{h V V} &= \kappa_V \times g_{h V V}^{\mathrm{SM}}~,
\end{aligned}
\end{equation}
with 
\begin{equation}
    \begin{aligned}
         \kappa_F &= \frac{\cos{\alpha}}{\sin{\beta}}~, \\
         \kappa_V &= \sin{\beta} \cos{\alpha}-\sqrt{\frac{8}{3}} \cos{\beta} \sin{\alpha}~,
    \end{aligned}
\end{equation}
which are sensitive to the current Higgs measurements. In particular, the GM model is arguably the simplest custodially symmetric model whose $\kappa$'s can be larger than unity.  Also, because one major contribution to the di-Higgs production is a box diagram with an inner top-loop, the modifications to the Yukawa couplings also have a large impact on this process.

Here we briefly comment on the decoupling limit of the GM model\footnote{We note that the model does not have the limit of alignment without decoupling.}, which is an important region for the global fit as the conclusive discovery of new physics has yet been made to date. The decoupling limit of the GM model is achieved when $v_\Delta\to0$ and $\mu_1\to0$, as a result of which we have
\begin{equation}
\cos{\beta} \to 0, \quad \alpha \to 0, \quad M_1^2 \gg v^2  \text{~and~} \quad M_2^2 \to 0 ~.
\end{equation}
In this limit, the scalar masses reduce to
\begin{equation}
\begin{aligned}
m_{H_5}^2 &\to -\frac{3}{2} \lambda_5 v^2 + M_1^2 + M_2^2~, \\
m_{H_3}^2 &\to -\frac{1}{2} \lambda_5 v^2 + M_1^2~, \\
m_{H_1}^2 &\to M_1^2 - \frac{1}{2} M_2^2~, \\
m_h^2 &\to 8 \lambda_1 v^2 ~, 
\end{aligned}
\end{equation}
where only $h$ remains at the electroweak scale and acts exactly like the SM Higgs boson. Additionally, the mass spectrum of the exotic Higgs bosons satisfies the relation
\begin{equation}\label{eq:mass hierarchy}
    2m_{H_1}^2=3m_{H_3}^2-m_{H_5}^2
    ~.
\end{equation}

We now discuss the theoretical constraints on the parameter space. We consider three different sets of constraints at the tree level: the vacuum stability or the bounded from below (BFB) condition, the perturbative unitarity condition, and the unique vacuum condition\footnote{We remark that the theoretical bounds implemented in this work are conservative. The loop corrections may break these constraints~\cite{Chiang:2018cgb}. Because of the attention on LHC constraints, we use more relaxed bounds on the theory side.}.

The BFB condition ensures that there is a stable vacuum in the potential. As noted in Ref.~\cite{Hartling:2014zca}, the BFB constraint can be satisfied as long as the quartic terms of the scalar potential remain positive for all possible field configurations, and can be guaranteed by satisfying the following conditions:
\begin{equation}
\begin{array}{l}
\lambda_{1}>0 ~, \\
\lambda_{2}>\left\{\begin{array}{l}
-\frac{1}{3} \lambda_{3} \text { for } \lambda_{3} \geq 0 ~, \\
-\lambda_{3} \quad \text { for } \lambda_{3}<0 ~,
\end{array}\right. \\
\lambda_{4}>\left\{\begin{array}{l}
-\frac{1}{2} \lambda_{5}-2 \sqrt{\lambda_{1}\left(\frac{1}{3} \lambda_{3}+\lambda_{2}\right)} \quad \text { for } \lambda_{5} < 0 \text { and } \lambda_{3} \geq 0 ~, \\
-\omega_{+}(\zeta) \lambda_{5}-2 \sqrt{\lambda_{1}\left(\zeta \lambda_{3}+\lambda_{2}\right)} \text { for } \lambda_{5} < 0 \text { and } \lambda_{3}<0 ~, \\
-\omega_{-}(\zeta) \lambda_{5}-2 \sqrt{\lambda_{1}\left(\zeta \lambda_{3}+\lambda_{2}\right)} \text { for } \lambda_{5} \geq 0 ~,
\end{array}\right.
\end{array}
\end{equation}
where $\omega \in\left[\omega_{-}, \omega_{+}\right]$, and
\begin{equation}
    \omega_{\pm}(\zeta)=\frac{1}{6}(1-B) \pm \frac{\sqrt{2}}{3}\left[(1-B)\left(\frac{1}{2}+B\right)\right]^{1/2} ~,
\end{equation}
with
\begin{equation}
B \equiv \sqrt{\frac{3}{2}\left(\zeta-\frac{1}{3}\right)} \in[0,1], \text{  and  } \zeta \in [\frac{1}{3},1] ~.
\end{equation}

The perturbative unitarity condition requires that the largest zeroth partial-wave mode of all $2\to2$ scattering channels be smaller than $1/2$ at high energies. Such constraints of the GM model were first studied in Ref.~\cite{Aoki:2007ah} and shown to be
\begin{equation}
\begin{aligned}
&\left|6 \lambda_{1}+7 \lambda_{3}+11 \lambda_{2}\right| \pm \sqrt{\left(6 \lambda_{1}-7 \lambda_{3}-11 \lambda_{2}\right)^{2}+36 \lambda_{4}^{2}}<4 \pi ~, \\
&\left|2 \lambda_{1}-\lambda_{3}+2 \lambda_{2}\right| \pm \sqrt{\left(2 \lambda_{1}+\lambda_{3}-2 \lambda_{2}\right)^{2}+\lambda_{5}^{2}}<4 \pi ~, \\
&\left|\lambda_{4}+\lambda_{5}\right|<2 \pi, \quad\left|2 \lambda_{3}+\lambda_{2}\right|<\pi ~,
\end{aligned}
\end{equation}
in the high-energy limit.

The unique vacuum condition~\cite{Hartling:2014zca} requires that there be no alternative global minimum in the scalar potential to the custodially-conserving vacuum. To examine this condition, we first parametrize the triplet fields as
\begin{equation}
\operatorname{Re} \chi^{0}=\frac{1}{\sqrt{2}} \sin \theta, \quad \xi^{0}=\cos \theta ~,
\end{equation}
where $\theta \in [-\pi, \pi]$. Then, we scan over the $\theta$ interval and check whether there is a deeper point in the potential than the custodially-conserving limit lying at $\theta=\pi/4$.

\section{Global Fitting and Experimental Constraints}
\label{sec:Constraints}

In our global fits in the GM model, we utilize the \texttt{HEPfit} package which is based upon a Bayesian statistics approach.  The Bayes theorem states that
\begin{equation}
p(\vec{p} \mid \vec{d}, m)=\frac{p(\vec{d} \mid \vec{p}, m) \times p(\vec{p} \mid m)}{p(\vec{d} \mid m)} ~,
\end{equation}
where $p(\vec{d} \mid \vec{p}, m)$ is the likelihood, $p(\vec{p} \mid m)$ is the prior\footnote{Conceptually, a prior can either merely specifies the pre-knowledge of the parameter distributions, or further embed the behavior of the model. For example, the tadpole conditions given in Eq.~(\ref{eq:tadpole conditions}) can have non-physical solutions, such as duplicate vacua or imaginary VEVs (note that we have chosen the phase convention such that $v_\phi,v_\Delta$ are both real and positive). The exclusion of such data points can either be thought of as part of the prior or as part of the likelihood. In this work, we choose to interpret this in the former way, and consequently, the likelihood contains only the theoretical and experimental constraints.}, and $p(\vec{p} \mid \vec{d},m)$ is the posterior. These probability distributions are described by the model parameters $\vec{p}$, the data $\vec{d}$, and the prior knowledge $m$, which is defined by the mean values and variances of the input parameters. Thus, in addition to the experimental data that determine the likelihood, a prior that specifies the \textit{a priori} distributions of the model parameters is also required, in which we can freely embed our pre-knowledge of the model. Based on the posterior probability, we sample the restricted parameter
space and attribute the allowed parameter ranges with different confidence levels.  A confidence level (C.L.) is the percentage of all possible samples that is expected to include the true parameters.

As alluded to earlier, a similar global fit had been performed in Ref.~\cite{Chiang:2018cgb}. This work differs from it in the following ways. First, the theoretical constraints are refined according to Ref.~\cite{Aoki:2007ah} (as we discussed in Sec.~\ref{sec:The Georgi-Machacek Model}) and the experimental data are updated.  Second, we focus on the parameter space where the exotic Higgs masses are reachable according to the LHC sensitivity.  Finally, we change our scheme for the input parameters to achieve stabler numerical manipulations. We now address the details of the global fit.

\subsection{Prior choices and mass constraints}

In a typical Bayesian fit, it is important to select a reasonable prior, lest the fit leads to unwanted statistical biases or non-physical results, while at the same time embedding our pre-understanding of the model into the fit.
In our work, we choose the following seven potential parameters: $\lambda_2$, $\lambda_3$, $\lambda_4$, $\lambda_5$, $\mu_1$, $\mu_2$ and $m_2^2$ as the input parameters. We make this change compared to Ref.~\cite{Chiang:2018cgb} because of the limited precision-handling capability of computers, which could cause the inference of quartic couplings from the physical masses and VEVs to suffer from serious propagation of errors.  This is especially important to our fit as all of the theoretical constraints are imposed on the dimensionless parameters, which renders a relatively high demand of numerical precision.

We choose the priors of the dimensionless parameters to be uniform within the bounds specified by the perturbative unitarity conditions~\cite{Hartling:2014zca}. As for the other couplings, we choose to make them Gaussian-distributed and, therefore, they are in general unbounded. Moreover, we choose the $m_2^2$ prior to be uniformly distributed in a logarithmic scale. Finally, because we only focus on the mass ranges probable at the near future LHC, we impose auxiliary single-sided Gaussian constraints on the heavy scalar masses. The summary of the prior choices is given in Table~\ref{tab:priors setting}.

\begin{table}
	\begin{tabular}{l|llll}
		\hline Parameters \qquad \qquad& Feature\qquad \qquad &Shape\qquad \qquad &Mean\qquad \qquad& Error/Range\\
		\hline\hline \textbf{Input} & & &Priors\\ \hline
		$m_2^2$ / $\mathrm{GeV}^2$ & $\log$ & Gaussian &$10^{2}$& $\left(10^{-4}, 10^{8}\right)$ \\
		$\lambda_2$ & linear & Uniform & -- &$\left(-\pi, \pi\right)$ \\
		$\lambda_3$ & linear & Uniform & -- &$\left(-\pi, \pi \right)$ \\
		$\lambda_4$ & linear & Uniform & -- &$\left(-\pi, \pi \right)$ \\
		$\lambda_5$ & linear & Uniform & -- &$\left(-3\pi, 3\pi \right)$ \\
		$\mu_1$ / $\mathrm{GeV}$ & linear & Gaussian &$0$& $\left(-5\times10^{3}, 5\times10^{3}\right)$ \\
		$\mu_2$ / $\mathrm{GeV}$ & linear & Gaussian &$0$& $\left(-5\times10^{3}, 5\times10^{3}\right)$ \\
		\hline \textbf{Auxiliary} &&& Priors \\ \hline
		$m_{H_{1,3,5}}$ / $\mathrm{GeV}$ & $\mathbb{R}_+$ & AsymGaussian &$10^{-2}$&$(0,10^3)$ \\
		\hline 
		\hline
	\end{tabular} 
	\caption{Input parameters and corresponding prior choices, as well as the auxiliary constraints on the new scalar masses in our global fit.}
	\label{tab:priors setting}
\end{table}

\subsection{Experimental data from the colliders}

We mainly consider data from the LHC Higgs signal strength measurements and exotic scalar searches as our experimental constraints, supplemented with a few data from Tevatron. Based upon those used Ref.~\cite{Chiang:2018cgb}, we update with the latest data.

We show in Table~\ref{tab:signalstrengthinputs} in Appendix~\ref{appendix:exp list} the current sensitivity of each individual channel for the Higgs signal strengths. The new data that we add are quoted from Refs.~\cite{ATLAS-CONF-2020-027,ATLAS-CONF-2020-045,Aad:2020jym,CMS-PAS-HIG-17-026,ATLAS:2020qcv,CMS-PAS-HIG-19-014}. We define $\hat{\sigma}$ to be the ratio of the smallest uncertainty of all individual measurements in one table cell of Table~\ref{tab:signalstrengthinputs} ($\sigma_{min}$) to the weight of the corresponding production channel ($w$)\footnote{ For example, the smallest uncertainty of the 13-TeV $gg\to h\to WW$ measurements is given by Ref.~\cite{ATLAS-CONF-2020-027}, which gives the signal strength $\mu=1.08^{+0.19}_{-0.18}$, and thus $\sigma_{min}=0.18$. The weight $w$ is $100\%$ in this case, and eventually we have $\hat{\sigma}=\sigma_{min}/w=0.18$. As such, the corresponding cell in Table~\ref{tab:signalstrengthinputs} is painted green according to the color scheme shown under the table. }. We then use $\hat{\sigma}$ to give an estimate on the current sensitivity of each individual channel. We remark that $\hat{\sigma}$ relies on the individual measurements instead of the combined ones, and thus this quantity is only intended to deliver a rough precision estimate for each channel.

The direct search data are listed in Appendix~\ref{appendix:exp list}, with the old data (Tables~\ref{tab:neutral heavy higgs with fermions},\ref{tab:neutral heavy higgs with bosons},\ref{tab:neutral heavy higgs with higgs} and \ref{tab:charged heavy higgs with charged scalars}) separated from the new ones (Tables~\ref{tab:new neutral direct searches} and \ref{tab:new charged direct searches}).

\subsection{Global fit results}

\begin{figure}[htb]
    \centering
    \includegraphics[width=1\textwidth]{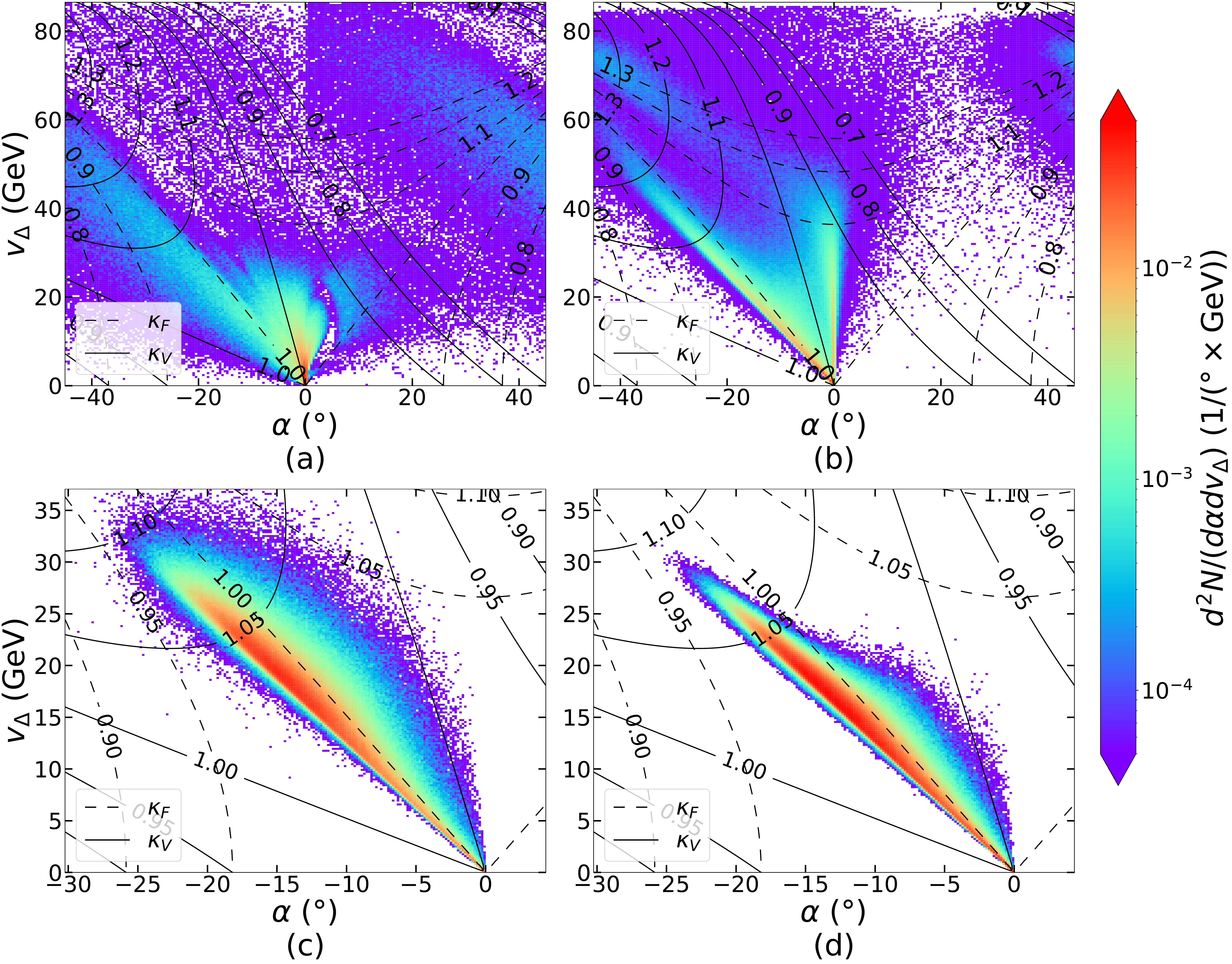}
    \caption{Normalized posterior distributions in the $\alpha$-$v_\Delta$ plane with (a) only the prior imposed, (b) theoretical constraints imposed, (c) theoretical constraints and Higgs signal strength constraints imposed, (d) theoretical constraints, Higgs signal strength constraints, and direct search constraints imposed. The dashed and solid curves represent the contours of $\kappa_{F}$ and $\kappa_{V}$, respectively.}
    \label{Fig:alpha-vDelta with different constraints}
\end{figure}

We show in Fig.~\ref{Fig:alpha-vDelta with different constraints} the results of the global fits, with different constraints imposed, in the $\alpha$-$v_\Delta$ plane. With our chosen prior, most of the data accumulate around the origin, which corresponds to the decoupling limit, as shown in Fig.~\ref{Fig:alpha-vDelta with different constraints}(a). 
After we impose the theoretical constraints, the data start to show a tendency towards the region around $\kappa_{F}\sim1$ in Fig.~\ref{Fig:alpha-vDelta with different constraints}(b). This is because the theoretical constraints tend to suppress the magnitudes of $\lambda_i$, which would in turn exclude the region where $M^2_{12}\to0$ or $M^2_{22}-M^2_{11}\gg1$ and thus cause the posterior of $\alpha$ to disfavor the point 0. Moreover, since the upper bound imposed on $m_{H_3}$ when $\alpha>0$, which is given by~\cite{Chiang:2015amq}
\begin{equation}
m_{H_{3}}^{2} \leq \frac{1}{2}\left(4 \lambda_{4}+\lambda_{5}\right) v^{2}+\sqrt{\frac{2}{3}} \frac{\sin \alpha \cos \alpha}{\sin \beta \cos \beta} m_{h}^{2} ~,
\end{equation}
is suppressed by the theoretical constraints, the $\alpha>0$ region is in tension with our prior setting that favors large exotic scalar masses. As a result, the region where $\kappa_F\sim1$, which has already been favored by the prior, becomes dominant in the posterior distribution.

Once the Higgs signal strength constraints are applied, the allowed phase space becomes apparently restricted, as shown in Fig.~\ref{Fig:alpha-vDelta with different constraints}(c).  The region around $\alpha\sim0$ becomes excluded because the signal strengths in the $WW$ and $ZZ$ channels are measured to be larger than the SM predictions, thus favoring the region where $\kappa_{V}>1.0$. Finally, in Fig.~\ref{Fig:alpha-vDelta with different constraints}(d), we observe that the direct search data further exclude more of the region where $\kappa_{V}>1.05$ and $\kappa_{F}>1.0$ because the data with larger $H_{3,5}^\pm$ branching ratios to certain channels, as discussed in Ref.~\cite{Chiang:2015amq}, are excluded by the experiments.

Before closing this section, we would like to add a remark on the $m_1^2$ parameter.  While $m_1^2$ has to be negative in the SM to generate a non-trivial vacuum, this is not necessary for the GM model, as the VEV in the $\phi_r$ direction can be induced by the interactions between $\Phi$ and $\Delta$. We find from the results of the global fit that $m_1^2$ is bounded from above at $\sim9000$~GeV$^2$. When $m_1^2$ increases, stronger interactions between the doublet and triplet fields are required to induce a VEV in the $\phi_r$ direction, and eventually this will be bounded by the theoretical constraints. This phenomenon is crucial to the discussion of EWPT in the next section.

\section{Electroweak Phase Transition and Gravitational Waves}
\label{sec:Electroweak Phase Transition and Gravitational Waves}

In this section, we discuss the EWPTs and the spectrum of induced GWs in the GM model. At high temperatures, thermal corrections dominate in the total potential and stabilize at the origin where the electroweak symmetry is preserved. When the temperature drops to a critical temperature $T_C$, where the potential develops another minimum of equal height to the origin, a non-trivial symmetry-breaking phase $\vec{h}(T=T_C)$ starts to form. If there exists a sufficiently high and wide potential barrier between the symmetric-phase vacuum and the broken-phase vacuum, then a first-order phase transition would take place. As the temperature further decreases, the potential barrier also lowers while the potential difference between the true and false vacua increases, eventually leading to bubble nucleation in the field plasma.  Collisions of these vacuum bubbles induce the production of stochastic GWs. In the following, we discuss the details of these dynamics in the GM model.

\subsection{Electroweak phase transitions}

\begin{figure}[htb]
    \centering
    \includegraphics[width=0.8\textwidth]{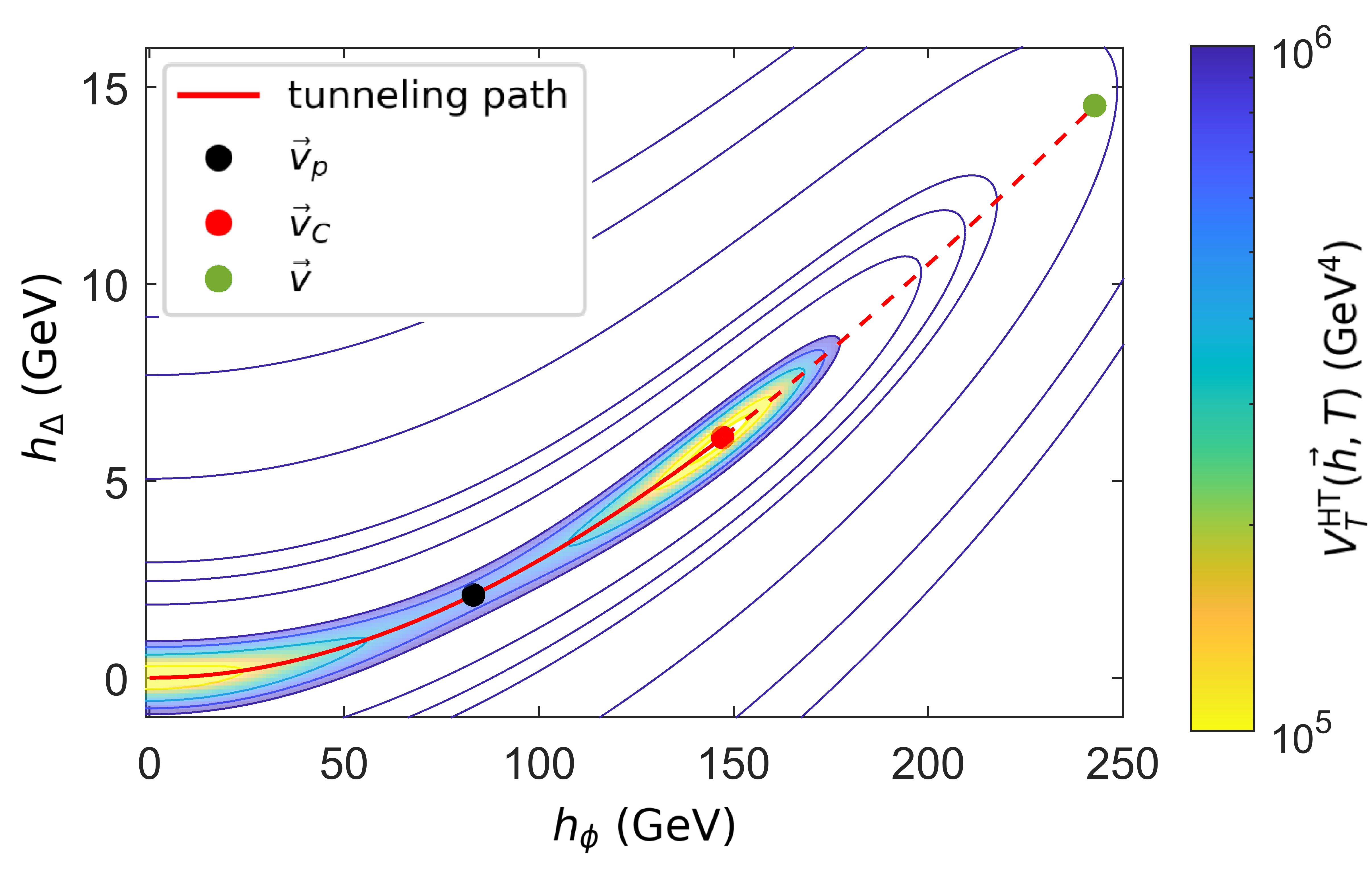}
    \caption{A schematic example of the first-order phase transition tunneling paths in the $(h_\phi$-$h_\Delta)$ plane.  The peak of the potential barrier at $T=T_C$ is denoted by $\vec{v}_p$, $\vec{v}_C$ represents the potential minimum at $T_C$, and $\vec{v}$ is the EW minimum at $T = 0$. The red solid curve represents the phase transition tunneling path, and the red dashed curve represents the extension of the tunneling path towards $\vec{v}$. The color map and the contours illustrate the potential distribution around the tunneling path at $T=T_C$.}
    \label{Fig:transition path}
\end{figure}

In our study, we assume that the EWPT takes place at a sufficiently high temperature such that the one-loop thermal corrections dominate over the Coleman-Weinberg potential, allowing an expansion of the thermal corrections to $\mathcal{O}\left(T^2\right)$. The overall potential at $T>0$ is then given by 
\begin{equation}\label{eq:HT potential}
V^{HT}_T(\vec{h}, T) = V_0(\vec{h}) + \frac{1}{2} \left( \Sigma_\phi h_\phi^2 + \Sigma_\chi h_\chi^2 +  \Sigma_\xi h_\xi^2\right) T^2 ~,
\end{equation}
where $V_0$ is the tree-level potential, $\vec h = (h_\phi, h_\xi, h\chi)$ and the thermal mass contributions
\begin{align}
\label{thermal masses contribution}
\begin{split}
\Sigma_{\phi}
&=
\frac{3 g^{2}}{16}+\frac{g^{\prime 2}}{16}+2\lambda_{1}+\frac{3 \lambda_{4}}{2}+\frac{1}{4} y_{t}^{2} \csc ^{2}{\beta} 
~, \\
\Sigma_{\chi}
&=
\frac{g^{2}}{2}+\frac{g^{\prime 2}}{4}+\frac{11 \lambda_{2}}{3}+\frac{7 \lambda_{3}}{3}+\frac{2\lambda_{4}}{3}
~,\\
\Sigma_{\xi}
&=
\frac{g^{2}}{2}+\frac{11 \lambda_{2}}{3}+\frac{7 \lambda_{3}}{3}+\frac{2\lambda_{4}}{3}
~,
\end{split}
\end{align}
with $y_t=\sqrt{2}m_t/v_\phi$ being the top Yukawa coupling, and $g$ and $g^\prime$ being respectively the $SU(2)_L$ and $U(1)_Y$ gauge couplings. Assuming that the custodial symmetry is still preserved at $T>0$, we set $h_\xi=h_\chi/\sqrt{2}=h_\Delta$. Obviously, the potential minimum approaches $\vec{v}=(v_\Phi, v_\Delta)$  as $T$ decreases. Fig.~\ref{Fig:transition path} shows a schematic example of the phase transition tunneling paths in the $h_\Delta-h_\phi$ plane in the GM model\footnote{While most benchmarks from our global fit give concave paths, there are also benchmarks that give either straight or convex paths.}. The thermal potential $\left(\Sigma_\phi h_\phi^2 + \Sigma_\chi h_\chi^2 + \Sigma_\xi h_\xi^2\right)T^2/2$, especially the $h_\phi^2$ term, is the primary source of the potential barriers, since it can lift the potential much higher than $V_0$ when $\vec{h}$ is small and $T$ is high.  On the other hand, $V_0$ plays the main role in determining the shape of the tunneling path, which is crucial to the phase transition characteristics.

We divide the EWPT calculation into two steps. First, we run a preselection to derive the critical VEVs ($\vec{v}_C$'s) and $T_C$'s of the data generated by \texttt{HEPfit} by numerically solving the equations $V_{T}^{HT}(\vec{v}_C, T_C)=V_{T}^{HT}(\vec{0}, T_C)$ and $\nabla V_{T}^{HT}=\vec{0}$. Since the preselection is just a simple procedure to pin down $v_C$'s and $T_C$'s of the data, we use \texttt{cosmoTransitions}~\cite{Wainwright:2011kj} to determine the order of the EWPT as well as to calculate the bubble dynamics.

To ensure the validity of the high-$T$ expansion, we focus on the data points with $T_C>60$~GeV. Roughly 10\% of the data points are found to generate strong first-order EWPTs. Among all the points generated by \texttt{HEPfit}, we have found no two-step EWPTs as claimed in Ref.~\cite{Zhou:2018zli}, which is partly due to the direct search data for the samples with larger $v_\Delta$ \footnote{Following the procedure outlined in Ref.~\cite{Zhou:2018zli}, we use \texttt{GMCalc v1.4.1}~\cite{Hartling:2014xma} with its default setting along with the constraints of the $S$ parameter, $b \to s \gamma$, and $B_s \to \mu^+ \mu^-$ to generate parameter samples.  Among such samples, \texttt{cosmoTransitions} finds that about $0.09\%$ gives rise to two-step phase transitions.  The smallest value of $v_\Delta$ in these samples is about $22.7$~GeV.  We have checked that they are all ruled out by direct search data, with some of the most constraining channels being $tt \to H_1 \to tt$~\cite{ATLAS:2018alq}, $VV \to H_5^{\pm\pm} \to W^\pm W^\pm [\to (\ell \nu) (\ell \nu)]$~\cite{CMS:2021wlt} and $b b \rightarrow H_{3}^{0} \rightarrow h Z \rightarrow(b b) Z$~\cite{ATLAS:2017xel}.} and partly due to the fact that our Bayesian scan fails to find those samples with smaller $v_\Delta$, particularly in the $v_\Delta \to 0$ limit as found in Ref.~\cite{Zhou:2018zli}, that could lead to two-step phase transitions.  This highlights how collider experiments can shed light on possible phase transition types of the model in the early Universe.

\begin{figure}[htb]
    \centering
    \includegraphics[width=0.7\textwidth]{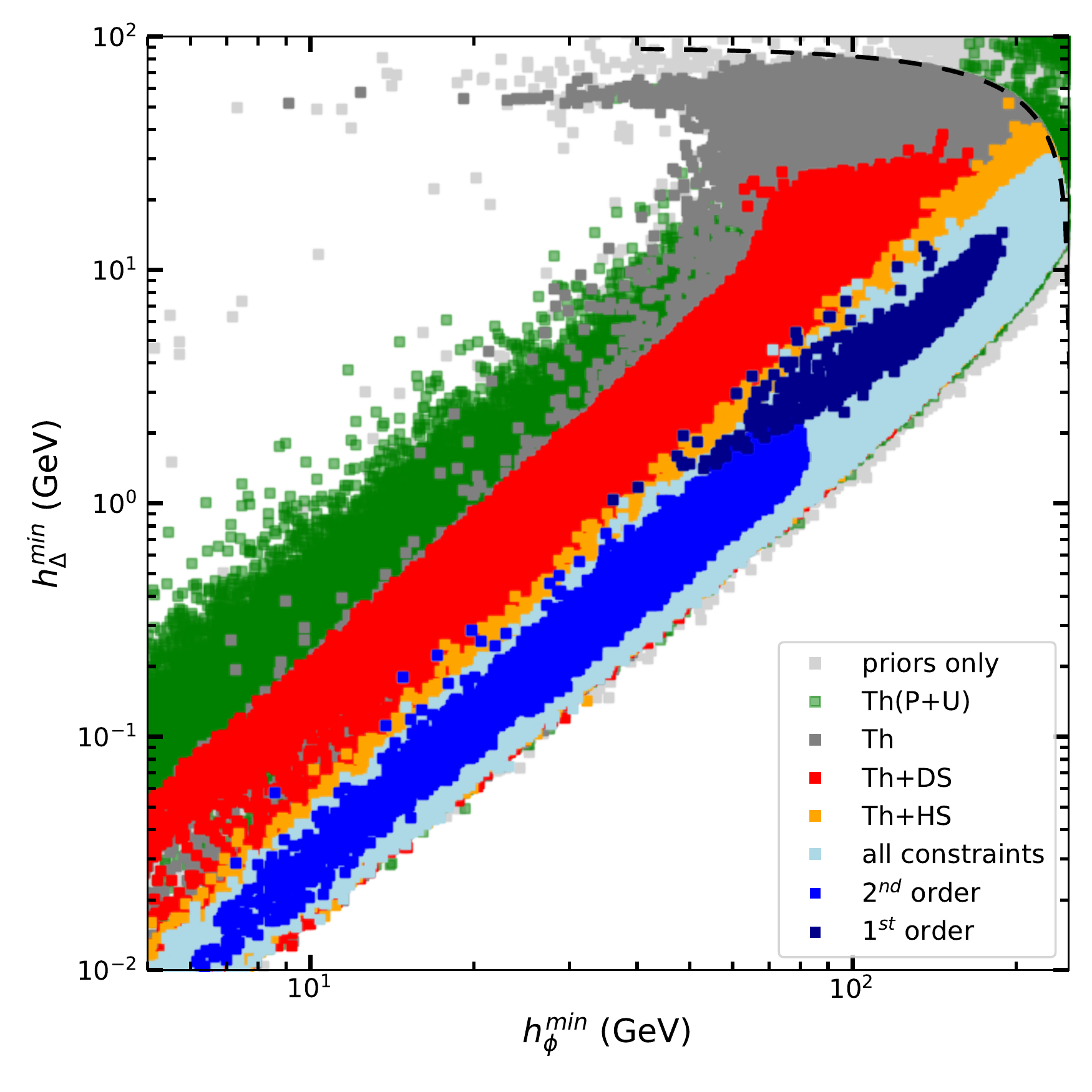}
    \caption{Scatter plot of $\vec{v}_C$ screened by the preselection method under different constraints. The light-gray, green, gray, red, orange and light blue points denote the data that pass the prior, the perturbative unitarity (P) and unique vacuum (U) constraints, the theoretical constraints, the theoretical and direct search constraints, the theoretical and Higgs signal strength constraints, and all of the above-mentioned constraints, respectively. The dark-blue first-order and blue second-order phase transition data points also pass all the constraints, and are further processed by \texttt{cosmoTransitions}, all with $T_C>60$~GeV. The black dashed curve denotes the contour of $|\vec{v}_{C}|=v$.}
    \label{Fig:transition points with different constraints}
\end{figure}

Fig.~\ref{Fig:transition points with different constraints} is a scatter plot of $\vec{v_C}$ calculated using the aforementioned preselection method under different constraints. We also present the $\vec{v_C}$  data that pass all the mentioned constraints and are further determined by \texttt{cosmoTransitions} to be of first-order and second-order phase transitions. After we impose the theoretical constraints, we observe that the BFB condition would exclude the data with $|\vec{v}_C| > v$ (the region to the right of the dashed curve), as can be seen by comparing the distributions of the green (perturbative unitarity and unique vacuum constraints imposed) and gray (all theoretical constraints imposed) data points. This is because the $V_0$'s of these excluded data points are not bounded from below when $|\vec{h}|\to\infty$, and hence the $V^{HT}_T$'s would create $\vec{v}_C$'s beyond $\vec{v}$ when $T$ increases. If we further impose either the Higgs signal strength or direct search constraint, the allowed range for $h_\Delta^{min}$ becomes even more restricted. The experimental constraints are thus responsible for the smaller $v_C$'s of the strong first-order EWPTs and the limitation on the values of $v_C/T_C$. This implies that the collider measurements are in fact good probes to the EWPT behavior of the GM model. We will illustrate this in more detail in Sec.~\ref{sec:Predictions}.

\begin{figure}[htb]
    \centering
    \includegraphics[width=0.85\textwidth]{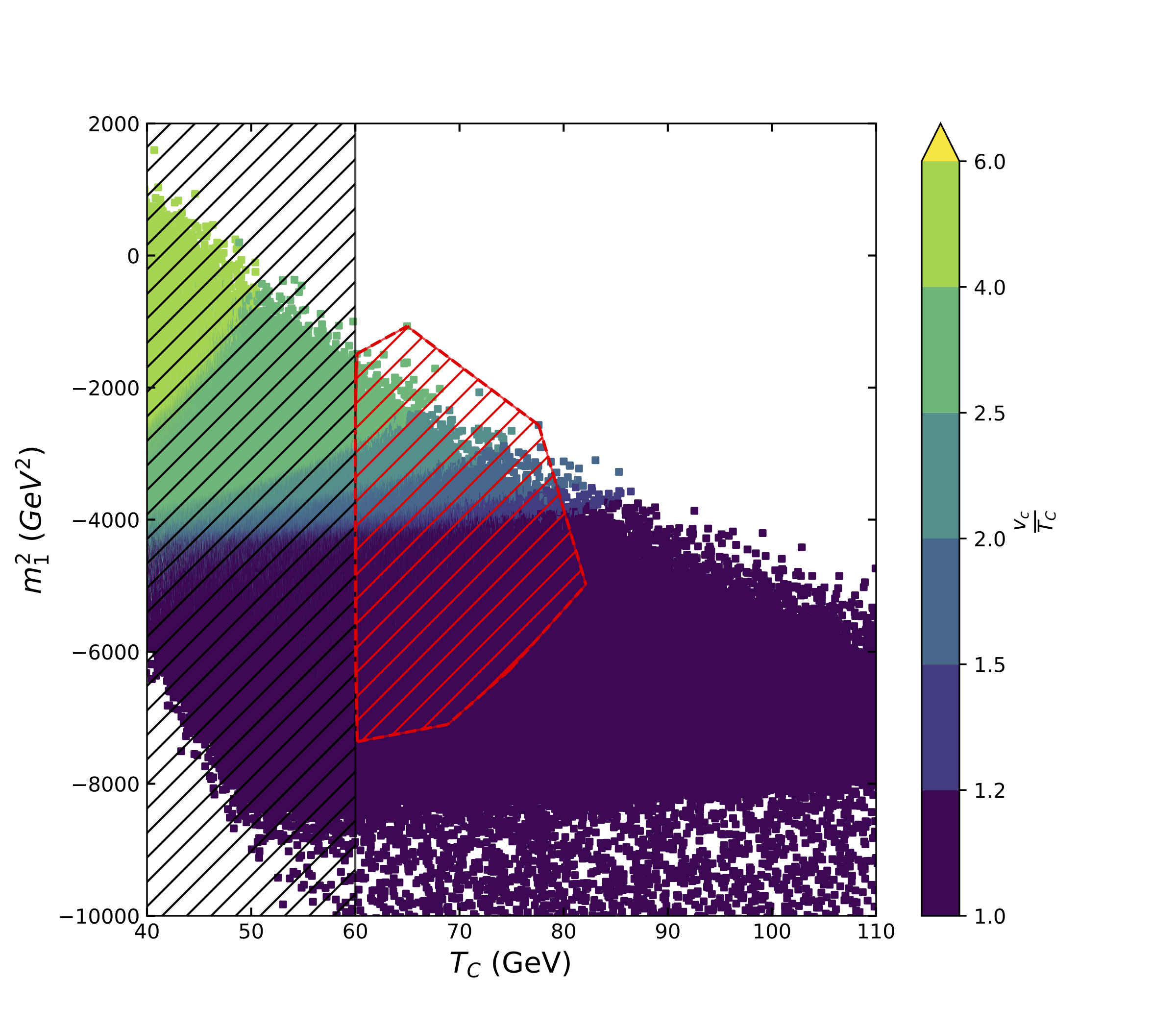}
    \caption{Scatter plot of the data points that pass all of the theoretical and experimental constraints in the $T_C$-$m_1^2$ plane. The color bar indicates the value of $v_C/T_C$. The first-order EWPT data points are contained in the red hatched region. We remark that the data points in the black hatched region violate the high-$T$ assumption. Also, $v_C$'s and $T_C$'s are derived using the preselection method, while the first-order EWPT data are further processed with \texttt{cosmoTransitions}.}
    \label{Fig:Tc-m1sq}
\end{figure}

We also illustrate the impact of the $m_1^2$ term in $V_0$ on $v_C$ and $v_C/T_C$ in Fig.~\ref{Fig:Tc-m1sq}. The black hatched region is first excluded because of the failure of high-$T$ expansion.  Some of the points falling within the red hatched region can give rise to first-order phase transitions.  As $m_1^2$ increases, $V_0$ becomes shallower in the $h_\phi$-direction, implying that the thermal corrections needed to lift the broken phase to the critical value are smaller, thus tending towards a lower $T_C$. Meanwhile, an increasing $m_1^2$ also lengthens the potential barrier and thus the transition path, which in turn enhances the phase transition strength $v_C/T_C$. Based on the same argument, we can see that as $m_1^2$ decreases, $T_C$ then tends to increase and $v_C/T_C$ tends to decrease. Consequently, as can be seen from the plot, most of the first-order phase transitions occur around $T_C\approx70$~GeV and are strong, with some of their $v_C/T_C$ reaching $2.5-4$ when $m_1^2\sim -2500$~GeV$^2$.

\subsection{Gravitational waves}

We now discuss the GWs induced from the bubble dynamics during EWPTs. The information of the stochastic GWs generated by the bubble dynamics of the strong first-order phase transitions can be completely accessed with two primary parameters: $\alpha_\mathrm{GW}$ and $\beta_\mathrm{GW}/H_n$~\cite{Kamionkowski:1993fg}. We adopt the model-independent methods from Refs.~\cite{Giese:2020rtr, Giese:2020znk, Guo:2021qcq}, which are based on the trace of the energy-momentum tensor, and define the strength parameter,
\begin{equation}
\alpha_\mathrm{GW}
=
\left.\frac{1}{3 \omega_{s}}\left[
T \frac{d \Delta V^{HT}_{T}(T)}{d T}
- \left(1+\frac{1}{c_{s}^{2}}\right) \Delta V^{HT}_{T}(T)
\right]\right|_{T=T_{n}}
~,
\end{equation}
where $w_s$ is the enthalpy density of hydrodynamics in the plasma outside the bubble (in the symmetry-preserving phase), $c_s$ is the speed of sound, and $\Delta V^{HT}_{T}(T) \equiv V^{HT}_{T}(\vec{h}(T), T) - V^{HT}_{T}(\vec{0}, T)$ is the potential difference between the broken phase and the symmetric phase at temperature $T$. $\alpha_\mathrm{GW}$ is related to the maximum available energy budget for GW emissions. Next, by assuming that the percolation takes place soon after the nucleation of the true vacua, which leads to the commonly used condition $T_{\ast} \simeq T_n$ where $T_\ast$ is the GW generation temperature and $T_n$ represents the nucleation temperature~\cite{Espinosa:2008kw,Ellis:2020awk}, $\beta_\mathrm{GW}/H_n$ is defined as
\begin{equation}\label{eq:beta_H}
\frac{\beta_{GW}}{H_{n}}=\left.T_{n} \frac{d}{dT} \left(\frac{S_{3}(T)}{T}\right)\right|_{T=T_{n}} ~,
\end{equation}
where $S_3$ denotes the three-dimensional on-shell Euclidean action of the instanton.  As $\beta_\mathrm{GW}/H_n$ is the inverse ratio of first-order EWPT duration to the universe expansion time scale. It defines the characteristic frequency of the GW spectrum produced from the phase transition.

\begin{figure}[htb]
    \centering
    \includegraphics[width=0.85\textwidth]{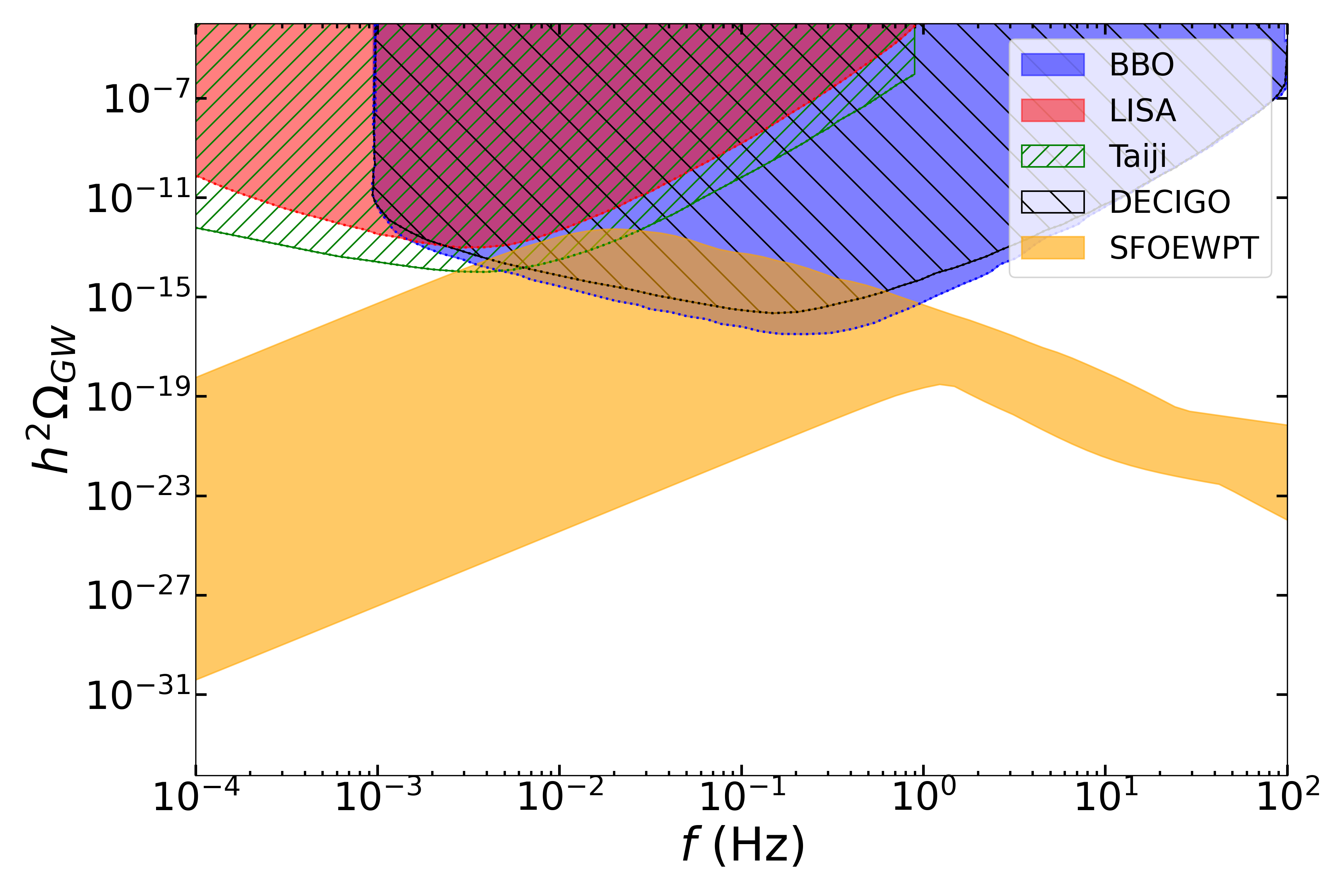}
    \caption{Spectra of GWs induced by the strong first-order EWPT data, as well as the power-law integrated sensitivities of various GW experiments.
    }
    \label{Fig:GW sensitivity}
\end{figure}

The main sources of the GWs generated during EWPTs are bubble collisions, sound waves, and turbulence, which have been well studied in the literature~\cite{Caprini:2015zlo,Cai:2017tmh}. According to the numerical estimations performed in Refs.~\cite{Huber:2008hg,Jinno:2016vai,Hindmarsh:2015qta,Caprini:2009yp,Espinosa:2010hh,Breitbach:2018ddu}, the GW spectra are given by
\begin{equation}\label{eq:GW amp}
    \begin{aligned}
h^{2} \Omega_{\mathrm{col}}(f) &=1.67 \times 10^{-5}\left(\frac{H_{n}}{\beta_\mathrm{GW}}\right)^{2}\left(\frac{\kappa_{\mathrm{col}} \alpha_\mathrm{GW}}{1+\alpha_\mathrm{GW}}\right)^{2}\left(\frac{100}{g_{*}}\right)^{\frac{1}{3}}\left(\frac{0.11 v_{w}^{3}}{0.42+v_{w}^{2}}\right) \frac{3.8\left(f / f_{\mathrm{col}}\right)^{2.8}}{1+2.8\left(f / f_{\mathrm{col}}\right)^{3.8}} ~,\\
h^{2} \Omega_{\mathrm{sw}}(f) &=2.65 \times 10^{-6}\left(\frac{H_{n}}{\beta_\mathrm{GW}}\right)\left(\frac{\kappa_{\mathrm{sw}} \alpha_\mathrm{GW}}{1+\alpha_\mathrm{GW}}\right)^{2}\left(\frac{100}{g_{*}}\right)^{\frac{1}{3}} v_{w}\left(\frac{f}{f_{\mathrm{sw}}}\right)^{3}\left(\frac{7}{4+3\left(f / f_{\mathrm{sw}}\right)^{2}}\right)^{7 / 2} ~,\\
h^{2} \Omega_{\mathrm{turb}}(f) &=3.35 \times 10^{-4}\left(\frac{H_{n}}{\beta_\mathrm{GW}}\right)\left(\frac{\kappa_{\mathrm{turb}} \alpha_\mathrm{GW}}{1+\alpha_\mathrm{GW}}\right)^{\frac{3}{2}}\left(\frac{100}{g_{*}}\right)^{1 / 3} v_{w} \frac{\left(\frac{f}{f_{\mathrm {turb}}}\right)^{3}}{\left(1+\frac{f}{f_{\mathrm{turb}}}\right)^{\frac{11}{3}} \left(1+\frac{8 \pi f}{H_{0}}\right)} ~,
    \end{aligned}
\end{equation}
where $g_{\ast}$ is the number of degrees of freedom at the domain wall decay time, which is $\approx 86$ in our study\footnote{The relativistic degrees of freedom for all the particles in the GM model in the early Universe is determined at $T=65$~GeV, which is the mean of $T_n$ in our studied samples.}. $\kappa_\mathrm{col}$, $\kappa_\mathrm{sw}$ and $\kappa_\mathrm{turb}$ are the transformation efficiencies of the first-order phase transition energy to kinetic energy, bulk motion of the fluid and turbulence, respectively, given by
\begin{equation}
\begin{aligned}
\kappa_\mathrm{col}&=\frac{1}{1+0.715 \alpha_\mathrm{GW}}\left[0.715 \alpha_\mathrm{GW}+\frac{4}{27} \sqrt{\frac{3 \alpha_\mathrm{GW}}{2}}\right]~,\\
\kappa_\mathrm{sw} &= \frac{\alpha_\mathrm{GW}}{0.73+0.083\sqrt{\alpha_\mathrm{GW}}+\alpha_\mathrm{GW}}~,\\
\kappa_\mathrm{turb} &= \xi_\mathrm{turb} \kappa_\mathrm{sw}
~,
\end{aligned}
\end{equation}
with the fraction of turbulent bulk motion  ($\xi_\mathrm{turb}$) assumed to be about $10\%$.
The red-shifted peak frequency of the GW spectra are given by
\begin{equation}\label{eq:GW freq}
    \begin{aligned}
f_{\mathrm{col}} &=16.5 \times 10^{-3}~ \mathrm{mHz} \times \left(\frac{0.62}{1.8-0.1 v_{w}+v_{w}^{2}}\right)\left(\frac{\beta_\mathrm{GW}}{H_{n}}\right)\left(\frac{T_{n}}{100 \mathrm{GeV}}\right)\left(\frac{g_{*}}{100}\right)^{\frac{1}{6}} ~,\\
f_{\mathrm{sw}} &=1.9 \times 10^{-2}~ \mathrm{mHz} \times \frac{1}{v_{w}}\left(\frac{\beta_\mathrm{GW}}{H_{n}}\right)\left(\frac{T_{n}}{100 \mathrm{GeV}}\right)\left(\frac{g_{*}}{100}\right)^{\frac{1}{6}} ~,\\
f_{\mathrm {turb}} &=2.7 \times 10^{-2}~ \mathrm{mHz} \times \frac{1}{v_{w}}\left(\frac{\beta_\mathrm{GW}}{H_{n}}\right)\left(\frac{T_{n}}{100 \mathrm{GeV}}\right)\left(\frac{g_{*}}{100}\right)^{\frac{1}{6}} ~,
    \end{aligned}
\end{equation}
where the bubble wall velocity $v_w\sim1$. 
Recent studies indicate that the contribution to the total GW spectrum from bubble collisions is negligible as very little energy is deposited in the bubble walls~\cite{Bodeker:2017cim}. In the following, we will restrict ourselves to the case of non-runaway bubbles, where the GWs can be effectively produced by the sound waves and turbulence.
Fig.~\ref{Fig:GW sensitivity} shows the GW spectra, represented by the yellow band based upon our two thousand data points, and the power-law integrated sensitivities of various GW experiments. We can see that the stronger the phase transition strength is, the larger the GW amplitude and the lower the peak frequency are.  This can be derived from Eqs~(\ref{eq:GW amp}) and (\ref{eq:GW freq}): when the phase transition strength is stronger, $T_C$ tends to be lower as implied in Fig.~\ref{Fig:Tc-m1sq}, and so does $T_n$, which leads to a larger $\alpha_\mathrm{GW} \propto T_n^{-2}$ and a smaller $\beta_\mathrm{GW}/H_n \propto  T_n$. The result shows that the GWs induced from the strong first-order EWPTs of the GM model can possibly be detected in \texttt{Taji}~\cite{Ruan:2018tsw}, \texttt{DECIGO}~\cite{Crowder:2005nr} and \texttt{BBO}~\cite{Sato:2017dkf} for $v_C/T_C\in[1,3.5]$, but not in \texttt{LISA}~\cite{2017arXiv170200786A}.

\begin{figure}[htb]
    \centering
    \includegraphics[width=0.8\textwidth]{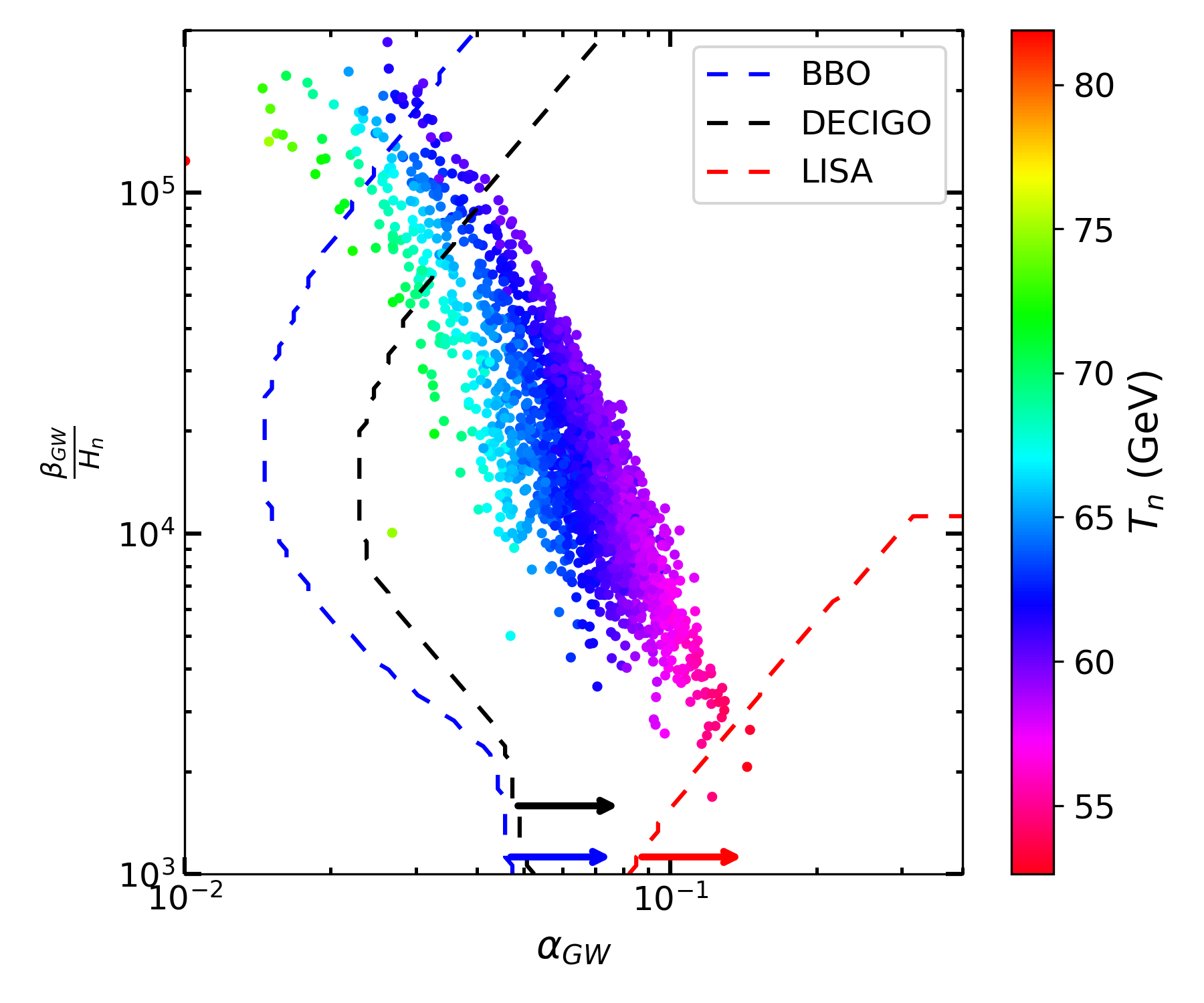}
    \caption{Scatter points of our data in the $\alpha_\mathrm{GW}$-$\beta_\mathrm{GW}/H_n$ plane. The color bar denotes $T_n$ of the data. The dashed curves represent the SNR thresholds of the listed GW experiments. The data to the right of the curves are detectable in the corresponding experiments.}
    \label{Fig:GW detectability}
\end{figure}

The detectability of the GW signals is evaluated by the corresponding signal-to-noise ratio (SNR)~\cite{Breitbach:2018ddu,Caprini:2015zlo}, given by
\begin{equation}
\rho=\sqrt{\mathcal{N} \mathcal{T}_{\mathrm{obs}} \int_{f_{\min }}^{f_{\max }} d f\left[\frac{h^{2} \Omega_{\mathrm{GW}}(f)}{h^{2} \Omega_\mathrm{exp }(f)}\right]^{2}} ~,
\end{equation} 
where $h^2\Omega_\mathrm{exp}$ is the effective noise energy density. $\mathcal{N}$ is the number of independent observatories of the experiment, which equals one for the auto-correlated experiments, and equals two for the cross-correlated experiments. $\mathcal{T}_\mathrm{obs}$ is the duration of the observation in units of year, assumed here to be four for each experiment as done in Ref.~\cite{Breitbach:2018ddu}. We summarize our assumptions and the features of interferometers in Table~\ref{tab:GW experiments}. We then extract the GW SNR thresholds assuming $T_n=65$~GeV from the documentations of the experiments. This temperature is chosen to be the same as the average value of $T_n$ for our strong first-order EWPT samples.  In Fig.~\ref{Fig:GW detectability}, the GW SNR thresholds are illustrated in the $\alpha_\mathrm{GW}$-$\beta_\mathrm{GW}/H_n$ plane, on which we also scatter our data. The data in the  regions to the right of the curves are above the SNR thresholds of the corresponding  GW observatories. As can be seen in the plot, most of our data are detectable and able to be separated from the instrumental noise in \texttt{BBO} and \texttt{DECIGO}.  The SNR threshold curves would have a small shift toward lower left if we choose a slightly higher $T_n$.

\begin{table}
\centering
\begin{tabular}{|l|l|c|c|c|c|}
\hline Experiment & Frequency range & $\rho_{\mathrm{thr }}$ & $\mathcal{N}$ & $\mathcal{T}_{\mathrm{obs }}[\mathrm{yrs}]$ & Refs. \\
\hline LISA & $10^{-5}-1 \mathrm{~Hz}$ & 10 & 1 & 4 & \cite{LISA:2017pwj,Robson:2018ifk} \\
DECIGO & $10^{-3}-10^{2} \mathrm{~Hz}$ & 10 & 2 & 4 & \cite{Sato:2017dkf,Yagi:2011yu,Yagi:2013du} \\
BBO & $10^{-3}-10^{2} \mathrm{~Hz}$ & 10 & 2 & 4 & \cite{Crowder:2005nr,Yagi:2011yu,Yagi:2013du} \\
\hline
\end{tabular}
\caption{Summary of the parameters and assumptions used for the projected space-based interferometers. \texttt{Taji} is not listed here because it does not release the effective noise energy density.}
\label{tab:GW experiments}
\end{table}

\section{Predictions}
\label{sec:Predictions}

In this section, we summarize and predict some of the most important and experimentally promising observables with our data, including those that are discussed in Section~\ref{sec:Constraints} and those that can further generate strong first-order EWPTs and GWs through bubble dynamics, as discussed in Section~\ref{sec:Electroweak Phase Transition and Gravitational Waves}. In the following plots, The former are presented with gray scatter points and the latter are shown with colored histograms.

\begin{figure}[htb]
    \centering
    \includegraphics[width=0.8\textwidth]{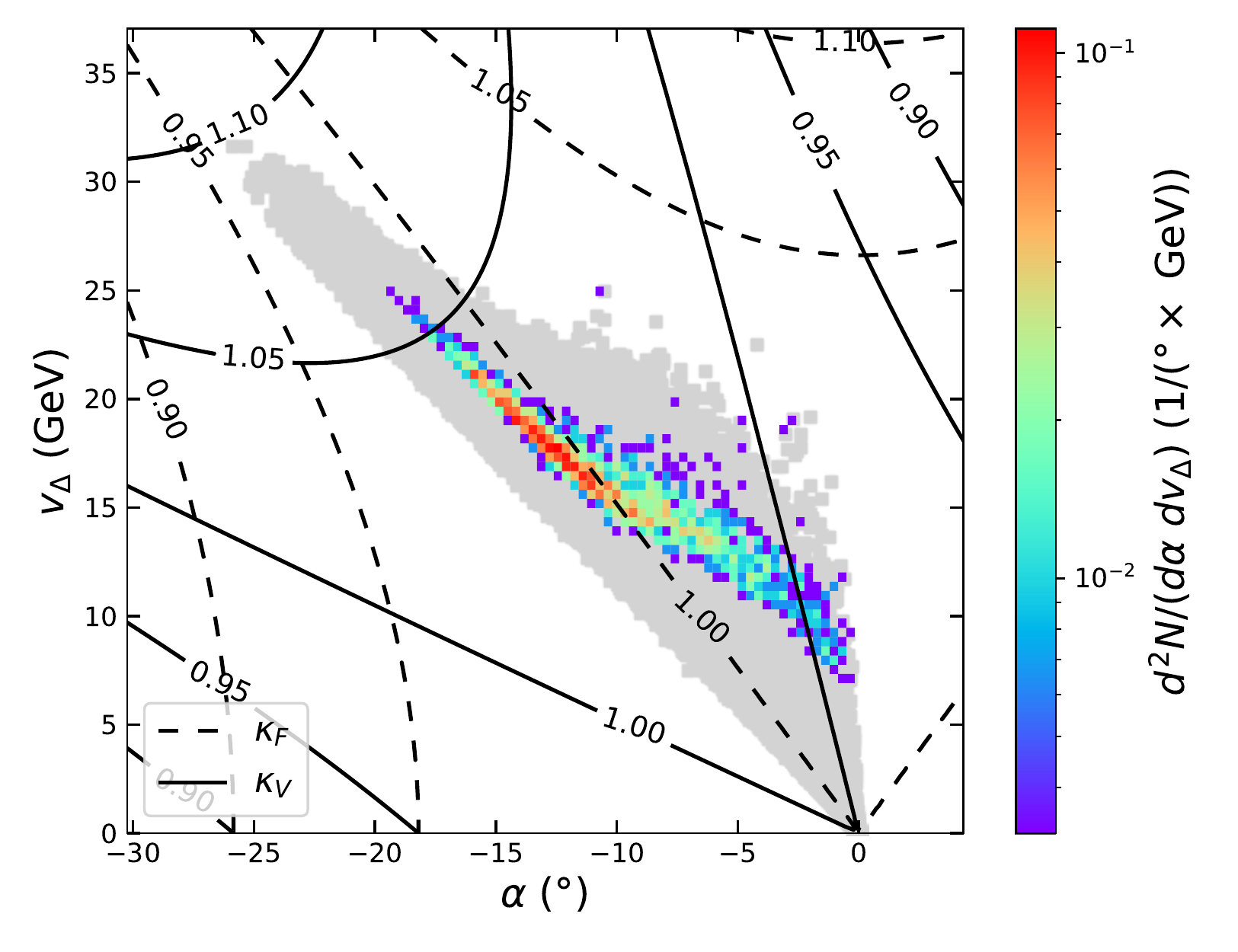}
    \caption{The predictions of our data in the $\alpha$-$v_\Delta$ plane. The gray scatter points represent the data that pass the \texttt{HEPfit}-level constraints, while the 2D colored histogram denotes the number density of the data that can further induce strong first-order EWPTs with $T_C>60$~GeV. The same plotting scheme is applied to all the following plots. The dashed curves and solid curves represent the contours of $\kappa_{F}$ and $\kappa_{V}$ respectively.}
    \label{Fig:alpha-vDelta with all constraints}
\end{figure}

Fig.~\ref{Fig:alpha-vDelta with all constraints} shows the prediction in the $\alpha$-$v_\Delta$ plane. The strong first-order phase transition data accumulate around $v_\Delta\in[15,20]$~GeV and $\alpha\in[-15^\circ,-10^\circ]$, corresponding to $\kappa_F\sim1$ and $\kappa_V\in(1.0,1.05)$, while they are mostly confined within $v_\Delta\in[5,25]$~GeV and $\alpha\in[-25^\circ,0^\circ]$. No data show up in the decoupling region because a SM-like potential could only induce a smooth crossover rather than strong first-order EWPTs.

\begin{figure}[htb]
    \centering
    \includegraphics[width=0.8\textwidth]{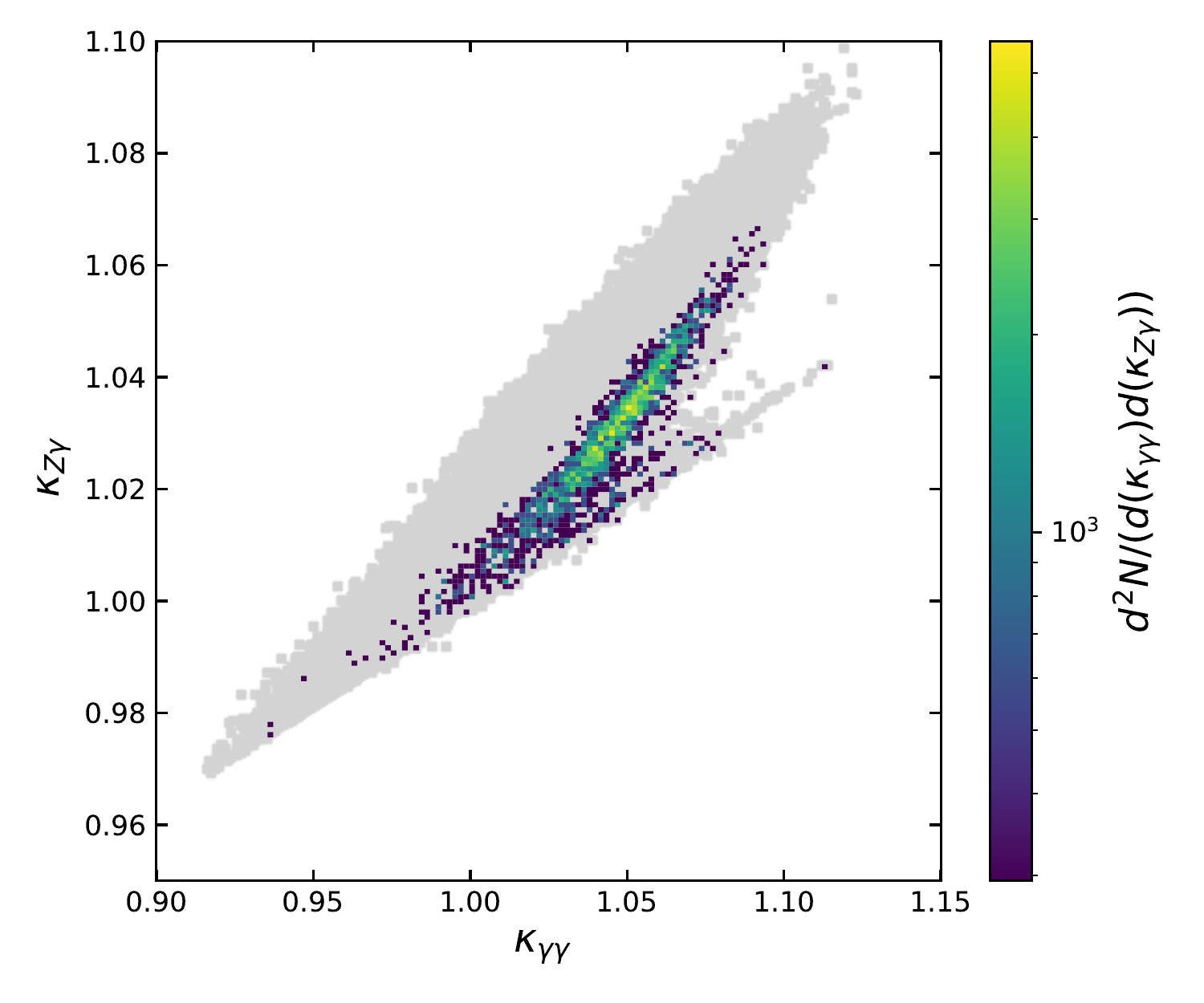}
    \caption{Prediction of our data in the $\kappa_{\gamma\gamma}$-$\kappa_{Z\gamma}$ plane.  The plotting scheme is the same as Fig.~\ref{Fig:alpha-vDelta with all constraints}.}
    \label{Fig:rh_gaga-rh_Zga}
\end{figure}

In Fig.~\ref{Fig:rh_gaga-rh_Zga}, we show the prediction in the $\kappa_{\gamma\gamma}$-$\kappa_{Z\gamma}$ plane, where $\kappa_{Z\gamma}$ and $\kappa_{\gamma\gamma}$ are the ratios of the loop-induced $hZ\gamma$ and $h\gamma\gamma$ couplings to the respective SM predictions. Compared to the result given in Ref.~\cite{Chiang:2018cgb}, we do not observe any data points around $\kappa_{Z\gamma}\sim0.1$ after imposing the Higgs signal strength constraints, and we find that it is ruled out by the new 13-TeV $h\to Z\gamma$ measurements~\cite{CMS-PAS-HIG-19-014,ATLAS:2020qcv}. Our results show that $hZ\gamma$ and $h\gamma\gamma$ couplings are positively correlated and, while most data give $\kappa_{Z\gamma}\sim 1.02$ and $\kappa_{\gamma\gamma}\sim 1.03$ at the \texttt{HEPfit}-level, the peak in the $\kappa_{\gamma\gamma}$-$\kappa_{Z\gamma}$ plane starts to approach $(1.03,1.05)$ after we require strong first-order EWPTs. Thus, a more precise measurement of these couplings can be a good probe to the EWPT behavior of the GM model.

\begin{figure}[h!]
    \centering
    \includegraphics[width=.47\linewidth]{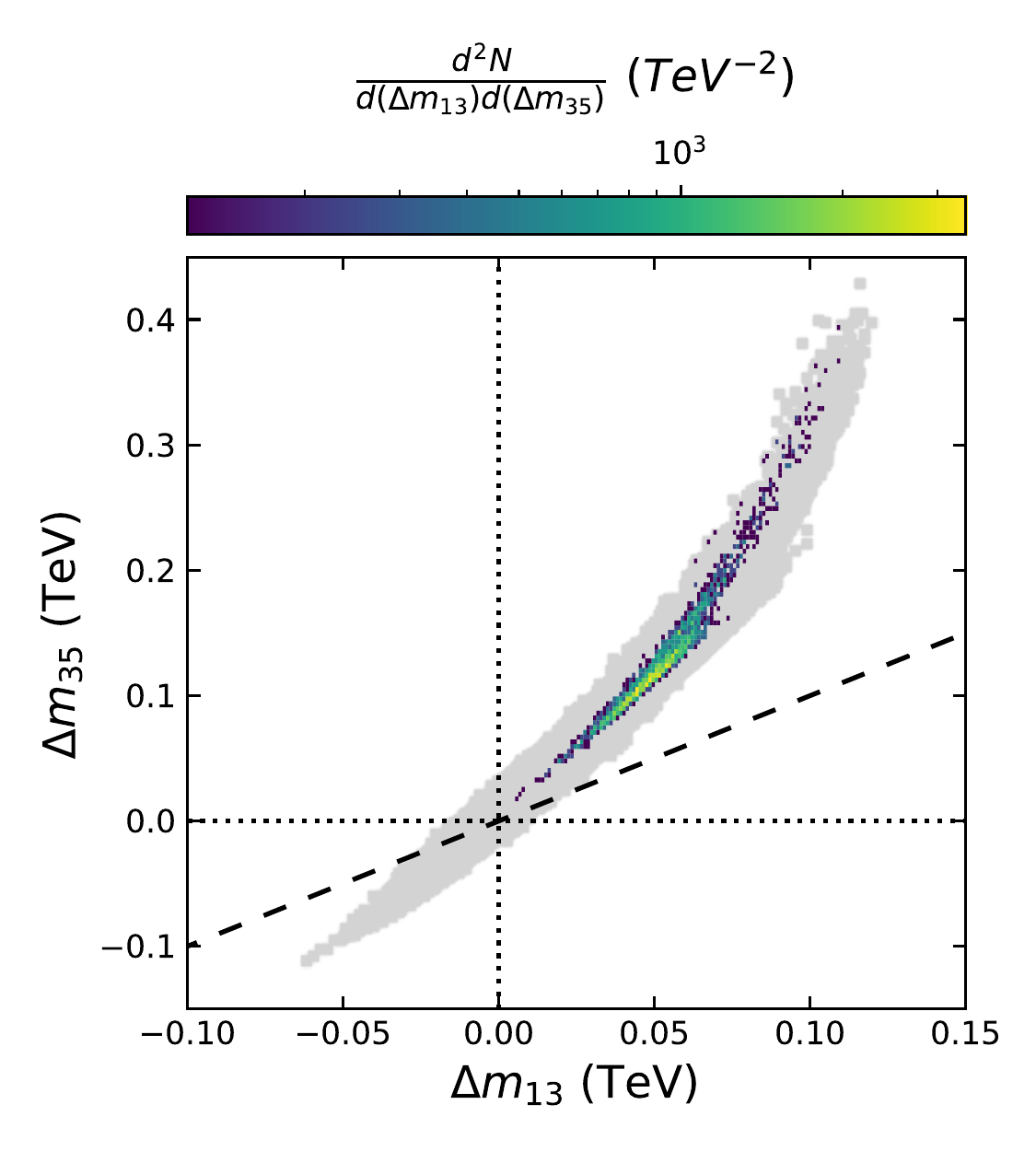}
    \includegraphics[width=.47\linewidth]{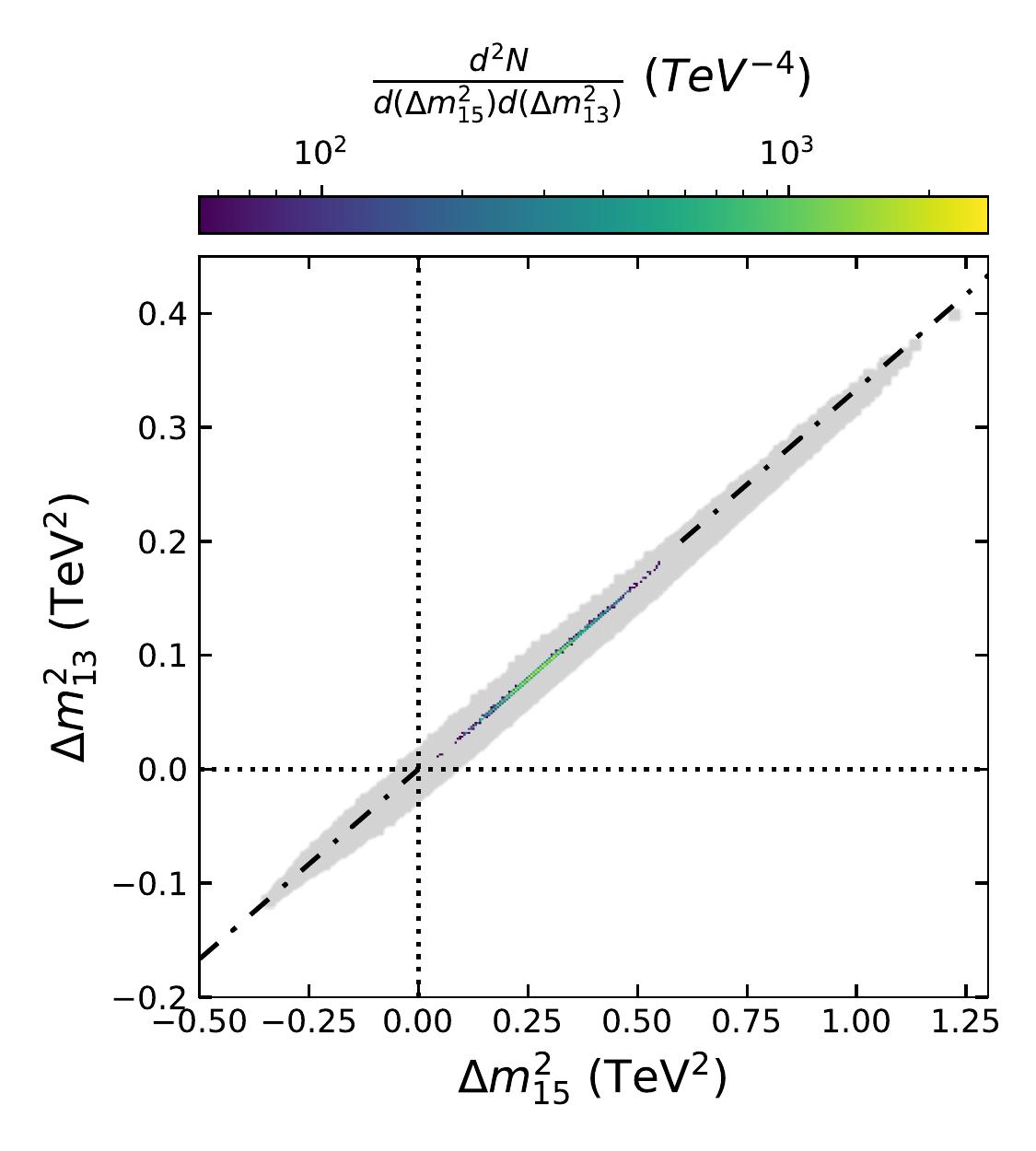}
    \\
	\vspace{-0.6cm}\hspace{1.cm}(a)
	\hspace{7cm} (b)
	\\
    \includegraphics[width=.47\linewidth]{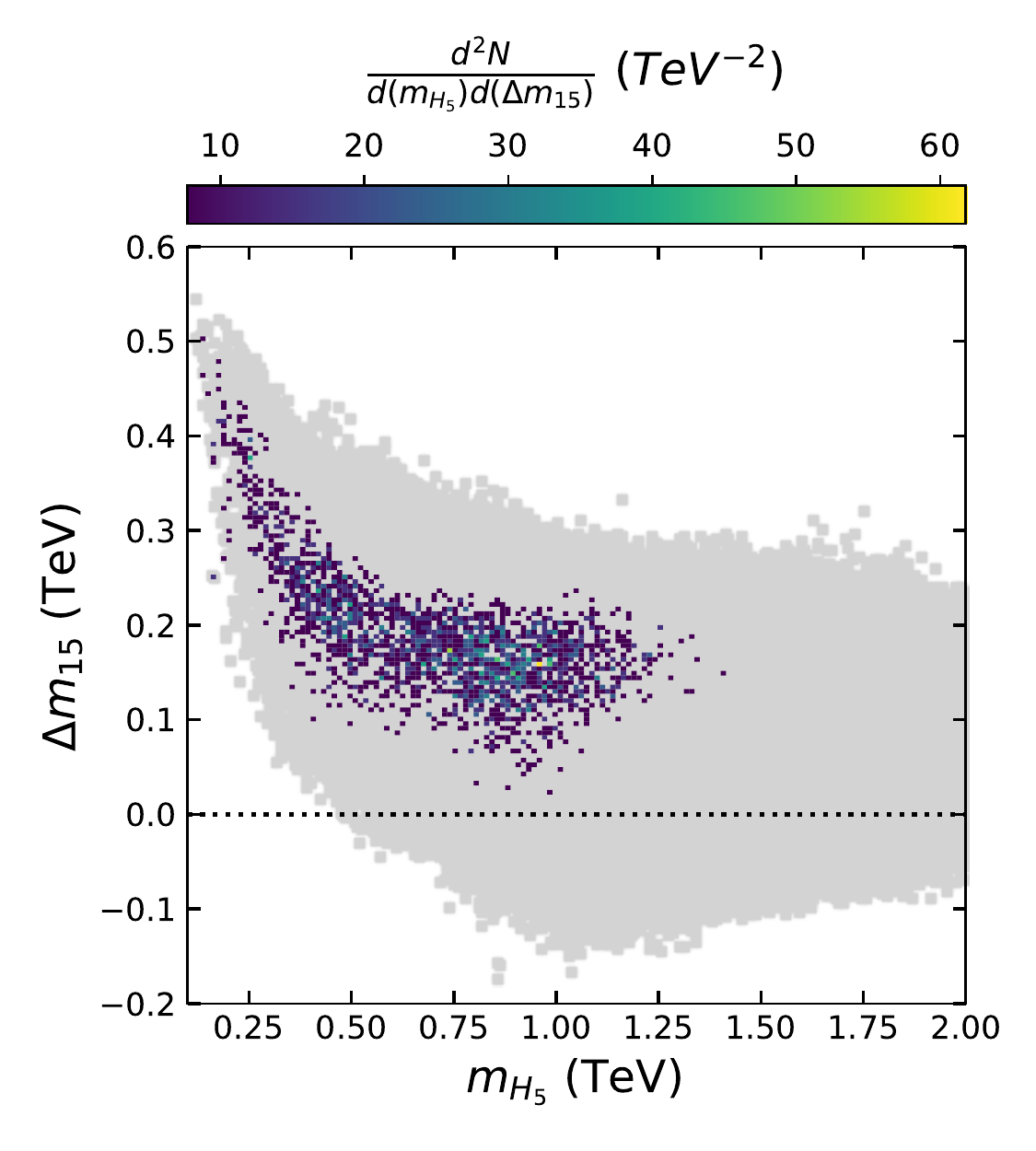}
    \\
	\vspace{-0.6cm} \hspace{.5cm}(c)
    \caption{Predictions of our data in (a) the $\Delta m_{13}$-$\Delta m_{35}$ plane, (b) the $\Delta m^2_{15}$-$\Delta m^2_{13}$ plane, and (c) the $m_{H_5}$-$\Delta m_{15}$ plane. The plotting scheme is the same as Fig.~\ref{Fig:alpha-vDelta with all constraints}. The slope of the dashed line in plot (a) is $1$, and that of the dot-dashed line in plot (b) is $1/3$.}
    \label{Fig:mass differences}
\end{figure}

Let's define the mass differences and the mass squared differences respectively as $\Delta m_{ij} \equiv m_{H_i}-m_{H_j}$ and $\Delta m_{ij}^2 \equiv m_{H_i}^2-m_{H_j}^2$, for $i,j=1,3,5$.  Fig.~\ref{Fig:mass differences} shows various mass relations according to our scan results.    As indicated by the gray points in Fig.~\ref{Fig:mass differences}(a), the constrained parameter space tends towards $\Delta m_{13}\in (-70,120)$~GeV, $\Delta m_{35} \in (-120,450)$~GeV, and $\Delta m_{15} \in (-200,550)$~GeV. We find that $\Delta m_{35}$ reaches its minimum when $m_{H_1} \sim 700$~GeV and its maximum when $m_{H_1} \sim 850$~GeV. After imposing the requirements of strong first-order EWPTs, all the data predict exclusively the mass hierarchy $m_{H_1} > m_{H_3} > m_{H_5}$, and most of them prefer a mass difference of around 50~GeV between $H_1$ and $H_3$ and around 100~GeV between $H_3$ and $H_5$. Such a mass hierarchy would limit certain scalar decay modes, such as $H^0_3 \to H^0_1 Z$, which has been searched for in the experiments $A_{13}^{\phi Z}$ and $A_{13b}^{\phi Z}$ as defined in Appendix~\ref{appendix:exp list}, and thus is another good probe to the EWPT behavior of the model.  With the auxiliary dashed line, one can also see that $\Delta m_{35}$ is always larger than $\Delta m_{13}$ for strong first-order EWPTs. Fig.~\ref{Fig:mass differences}(b) shows the distribution in the $\Delta m_{15}^2$-$\Delta m_{13}^2$ plane, to be compared with the mass relation predicted in the decoupling limit, given by Eq.~\eqref{eq:mass hierarchy} and indicated by the dot-dashed line.  Fig.~\ref{Fig:mass differences}(c) illustrates that the $H_5$ mass falls in the range of $[150,1500]$~GeV for the data points with strong first-order EWPTs.  When $m_{H_5} \lesssim 500$~GeV, there are possibilities that a larger mass gap exists between $H_1$ and $H_5$, making $m_{H_1}$ fall around $650$~GeV.

\begin{figure}[tb!]
    \centering
    \includegraphics[width=0.8\textwidth]{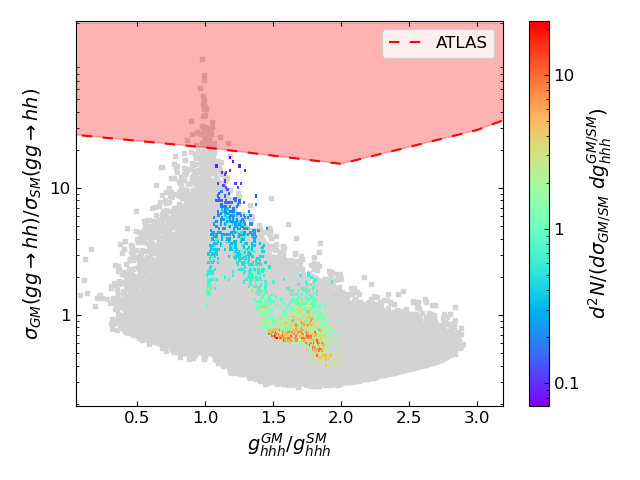}
    \caption{Predictions of the di-Higgs production cross sections against $g^{GM}_{hhh}/g^{SM}_{hhh}$, as well as the ATLAS bound at 95\% C.L.  The plotting scheme is the same as Fig.~\ref{Fig:alpha-vDelta with all constraints}.}
    \label{Fig:sigmagg sensitivity}
\end{figure}

The di-Higgs production cross sections are calculated with \texttt{Hpair}~\cite{Dawson:1998py} for the 13-TeV LHC collisions and illustrated in Fig.~\ref{Fig:sigmagg sensitivity}. At the leading order, the two triangle diagrams mediated by $h$ and $H_1$, as well as the box diagram with $t$ running in the loop give the most dominant contributions.  We also show the current 95\% C.L. upper limit given by ATLAS~\cite{ATLAS-CONF-2018-043}\footnote{The latest CMS constraint~\cite{CMS-PAS-HIG-17-030} is looser than the ATLAS constraint.}. We observe that except for a small patch of the parameter space with $g^{GM}_{hhh}/g^{SM}_{hhh}\sim1.2$, most of our data survive the ATLAS constraint and correspond to $g^{GM}_{hhh}/g^{SM}_{hhh}\in[1.4, 2.0]$. We discover that the strong first-order EWPT data would always lie in the region where $g^{GM}_{hhh}/g^{SM}_{hhh}>1$, and there are two peaks in the relative cross section distributions. We find that the left peak has a larger relative cross section because the associated values of $m_{H_1}$ are lighter, while those for the right peak are heavier and lead to smaller relative cross sections. Though with a relatively small portion, the data points with $g^{GM}_{hhh}/g^{SM}_{hhh} \in (1.1,1.4)$ have the prediction that $\sigma_{GM}(gg \to hh) / \sigma_{SM}(gg \to hh) \sim {\cal O}(2-20)$, to which future experiments are sensitive.  However, most of our data points have $g^{GM}_{hhh}/g^{SM}_{hhh}$ above $\sim 1.5$ with $\sigma_{GM}(gg \to hh) / \sigma_{SM}(gg \to hh)$ being smaller than the current sensitivity by at least one order of magnitude.

\begin{figure}[p]
\centering
	\includegraphics[width=0.48\textwidth]{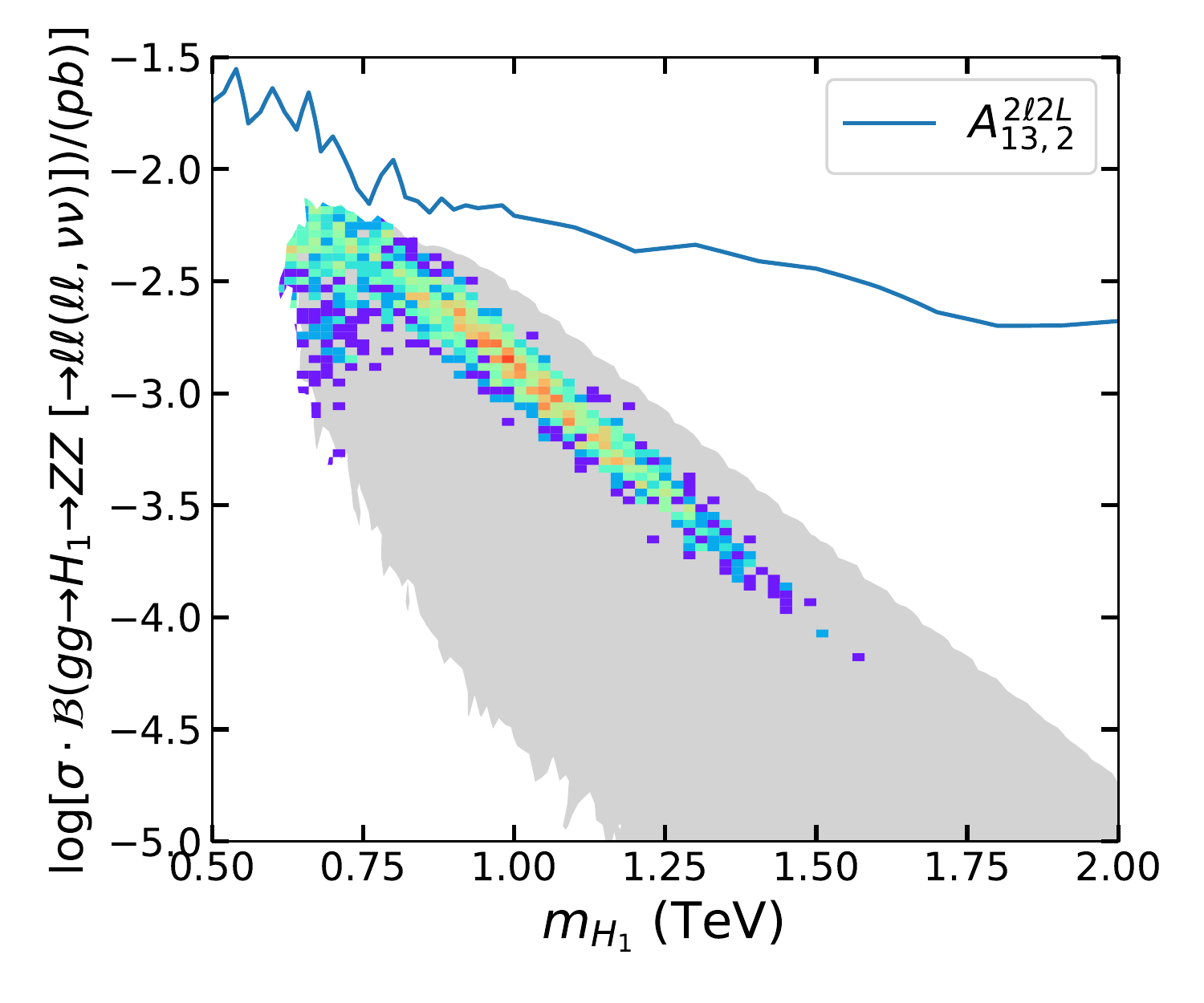}
	\includegraphics[width=0.48\textwidth]{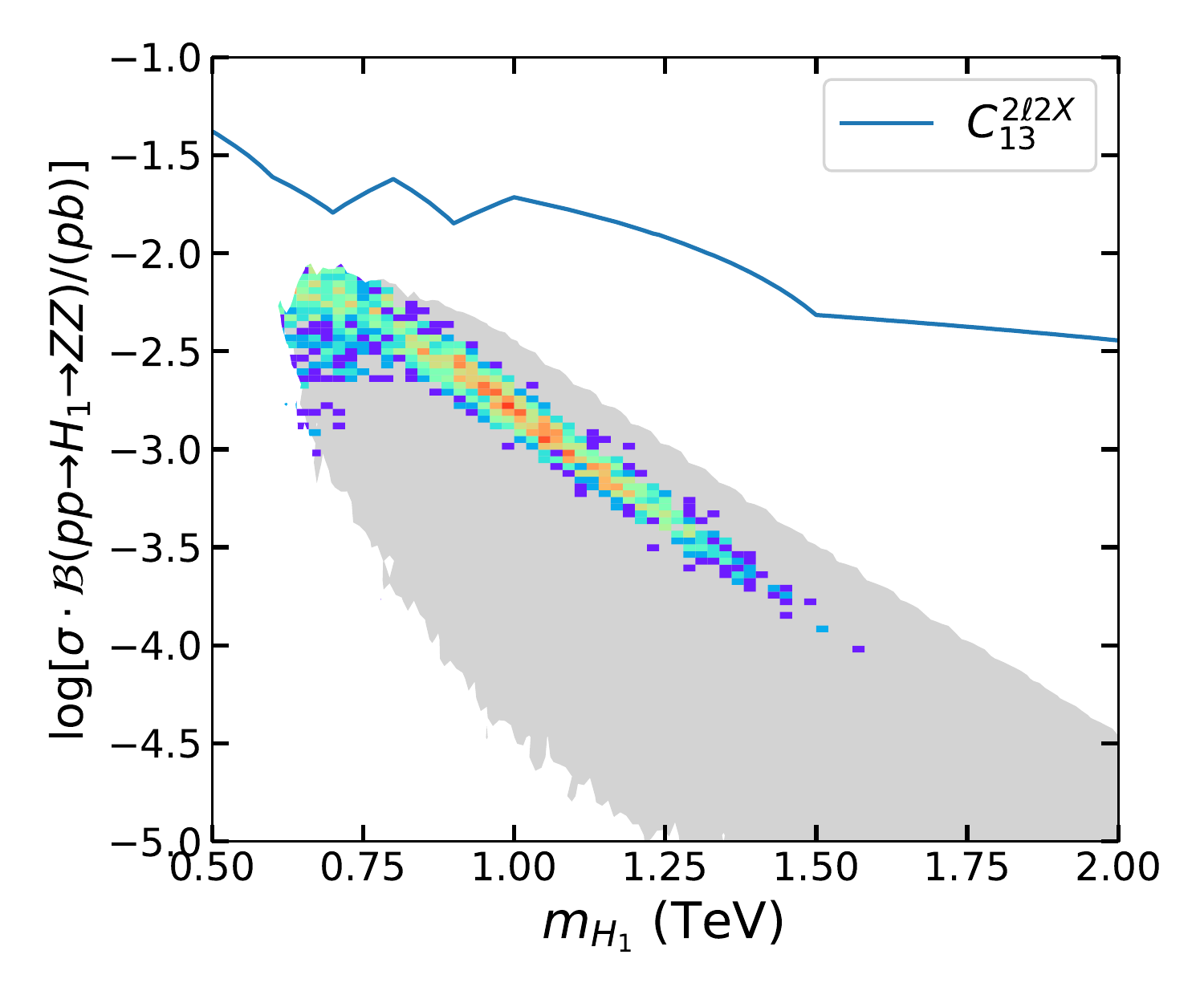}
	\\
	\vspace{-0.7cm} (a)
	\hspace{7.0cm} (b)
	\\
	\includegraphics[width=0.48\textwidth]{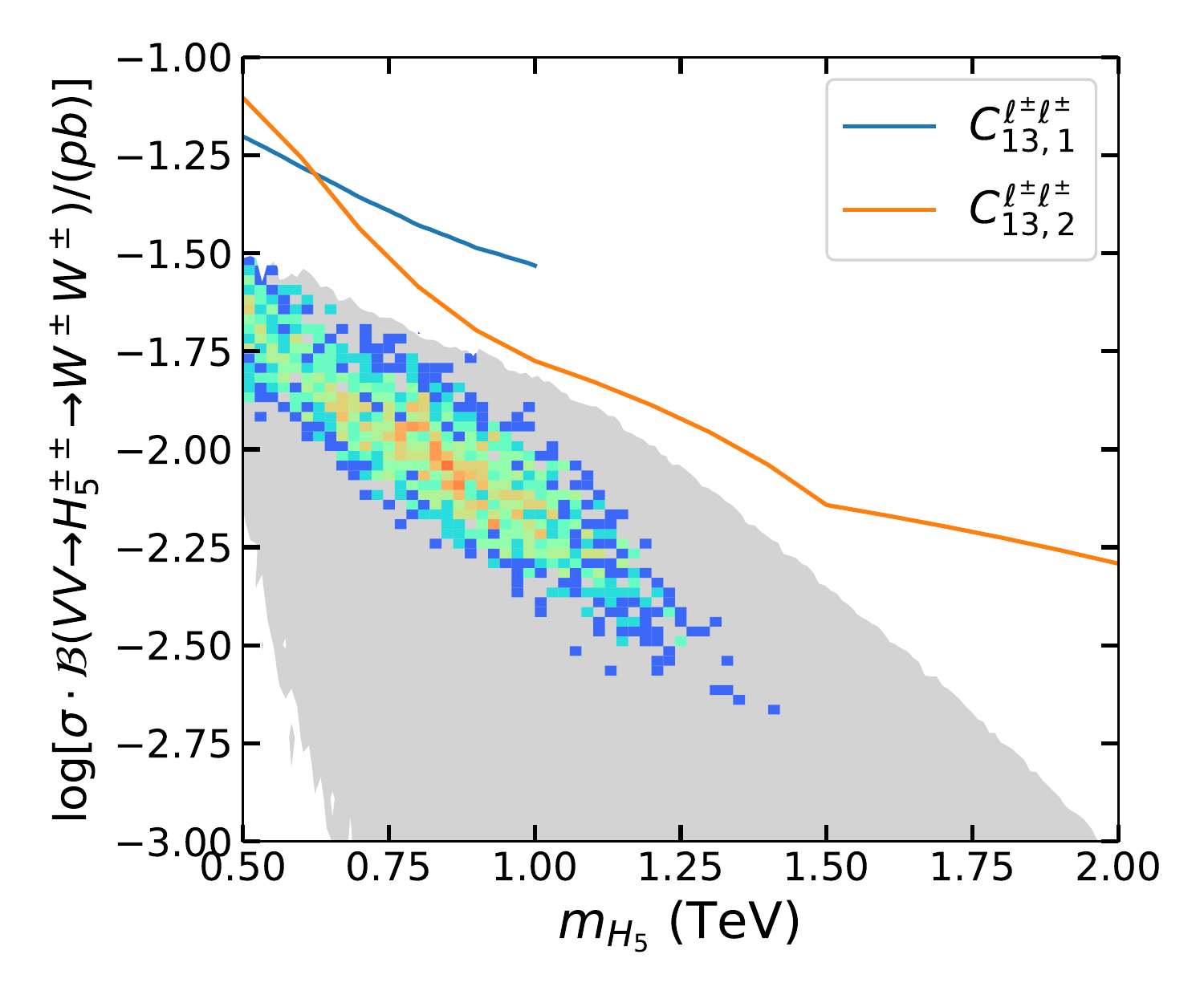}
	\includegraphics[width=0.48\textwidth]{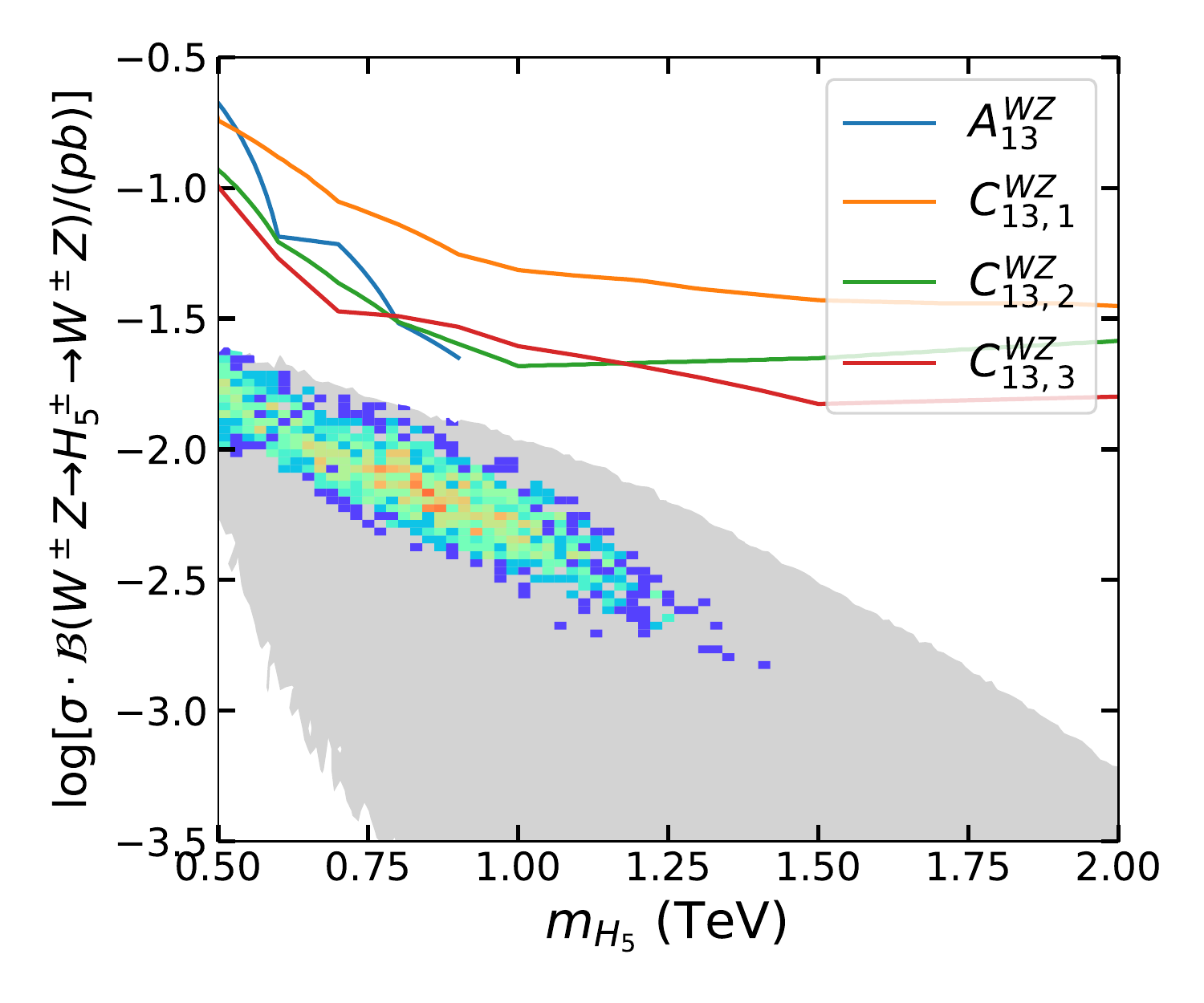}
	\\
	\vspace{-0.7cm} (c)
	\hspace{7cm} (d)
	\\
	\includegraphics[width=0.48\textwidth]{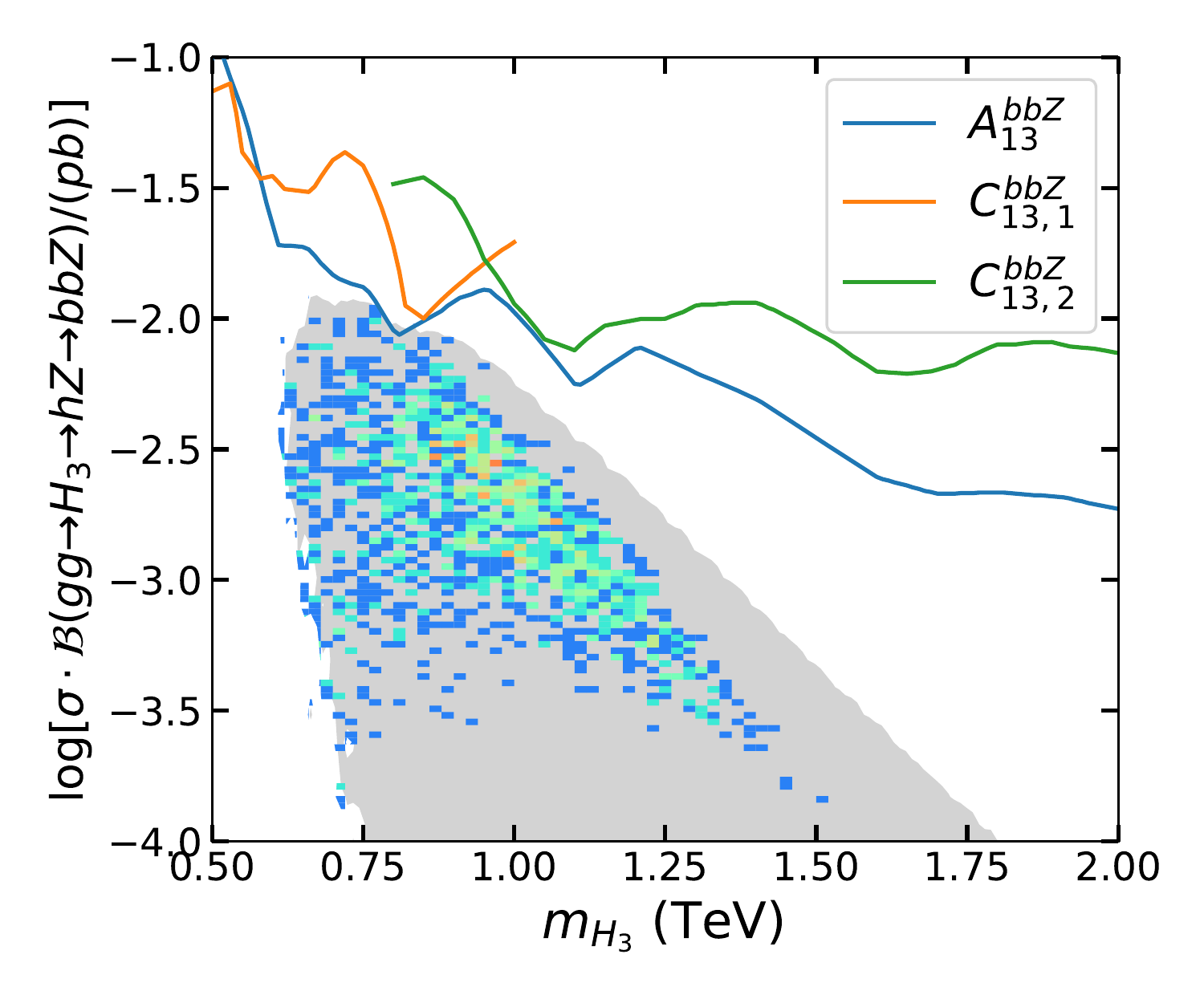}
	\hspace{-0.5cm}
	\includegraphics[width=0.08\textwidth]{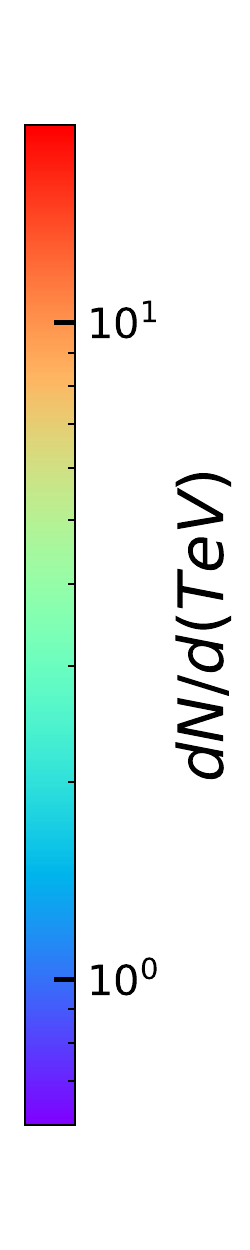}
	\\
	\vspace{-0.7cm} (e)
\caption{The predictions of the most constraining direct search channels. The colored curves indicate the 95\% C.L. limits imposed by the LHC measurements. The  plotting scheme is the same as Fig.~\ref{Fig:alpha-vDelta with all constraints}.}
\label{Fig:Direct Search constraints}
\end{figure}

Finally, we show some of the most constraining direct search channels for the GM model in Fig.~\ref{Fig:Direct Search constraints}. We also show the corresponding 95\% C.L. upper limits, which are given by ATLAS and CMS, including $A^{2\ell 2L}_{13,2}$, $C^{2\ell 2X}_{13}$, $C^{\ell^\pm \ell^\pm}_{13,1}$, $C^{\ell^\pm \ell^\pm}_{13,2}$, $A^{WZ}_{13}$, $C^{WZ}_{13,1}$, $A^{WZ}_{13,2}$, $A^{WZ}_{13,3}$, $A^{bbZ}_{13}$, $C^{bbZ}_{13,1}$ and $C^{bbZ}_{13,2}$. We remark that for each figure, the region below the gray area is not excluded, but is simply too unlikely to be sampled under the constraints imposed. From these results, we observe that the constraints from the $H_3^0$ channels are stronger than those from the $H^\pm_3$ channels, while the $H^\pm_5$ channels impose stronger constraints than the $H_5^0$ and $H_5^{\pm\pm}$ channels do. We can see that most of the mass ranges favored by the strong first-order EWPT data points are highly constrained, and thus these collider measurements also serve as good probes to the EWPT behavior of the GM model.

\section{Discussions and Summary}
\label{sec:Discussions and Summary}

We have performed global fits for the GM model with \texttt{HEPfit} to acquire the allowed phase space.  The considered constraints include theoretical bounds of vacuum stability, perturbative unitarity, and the unique vacuum, as well as experimental data of Higgs signal strengths and direct searches for exotic Higgs bosons.  We calculate $v_C$'s and $T_C$'s for the allowed phase space screened by \texttt{HEPfit} using the preselection method under the high-$T$ assumption, and then process the data with $v_C>60$~GeV by utilizing the \texttt{cosmoTransitions} package.  Based upon the scan results, we calculate the GW spectra induced by the bubble dynamics during the EWPT.

By comparing the results obtained at different levels of constraints in \texttt{HEPfit}, we demonstrate the tendency of each constraint level in the $\alpha$-$v_\Delta$ plane and identify the favored $\kappa_F$-$\kappa_V$ region. In particular, we find that there is an accumulation of data points around $(\kappa_F, \kappa_V)\sim(0.99,1.03)$ at all levels. In the vicinity of this point, $\kappa_F$ is almost one, and thus the cross sections of the ggF, bbH, and ttH production modes are nearly identical to the respective SM predictions. Moreover, since $\kappa_V\geq1$, the cross section of the VBF production mode would enhance, and so would the partial widths of the $h\to VV$ decays. Therefore, the $WW$ and $ZZ$ signal strengths are mostly enhanced within this region. We also study the previously unexplored region where $m_1^2>0$ and find that a nonzero $v_\phi$ can still be induced from the interactions between $\Phi$ and $\Delta$, although such a scenario is disfavored by the study of EWPT.

We find that the experimental constraints impose a relatively strong bound on the $\vec{v}_C$ distributions, especially in the $h_\Delta$ direction. Furthermore, we show the impact of $V_0$ on $T_C$ and $v_C/T_C$, especially regarding that $m_1^2$ has a major impact on the depth of the overall potential and thus on the EWPT characteristics.  

In the calculation of the induced GW spectra, we find that the peak frequency lies roughly within $[10^{-2},1]$~Hz and the corresponding amplitude $h^2 \Omega_{\text GW}$ can reach up to $10^{-12}$, which can be possibly detected by \texttt{Taji}, \texttt{DECIGO} or \texttt{BBO} in the near future, but not in \texttt{LISA}.

We calculate $\kappa_{Z\gamma}$ and find that the strong first-order EWPT phase space only affords a small deviation from the SM prediction. We also observe that the strong first-order EWPT data points all prefer the ``inverted'' mass hierarchy, $m_{H_1}>m_{H_3}>m_{H_5}$, with the masses lying within $[0.5,1.5]$~TeV.

Finally, we list some of the most constraining or physically interesting experiments, including the di-Higgs productions and several direct searches for exotic scalars. According to the \texttt{HEPfit} results, the di-Higgs production cross sections range from $0.3$ to $30$ times the SM prediction, and most data still lie below the sensitivity of the latest ATLAS measurement~\cite{ATLAS-CONF-2018-043}. The direct search channels we choose to show are $gg\to H_1 \to ZZ$, $pp \to H_1 \to ZZ$, $VV\to H^\pm_5 \to W^\pm W\pm$, $W^\pm Z \to H_5^\pm \to W^\pm Z$ and $gg\to H_3 \to h Z \to bbZ$ at $\sqrt{s}=13$ TeV, which serve as the most promising probes to the GM model in the near future LHC experiments.

\section*{Acknowledgments}
The authors would like to thank Otto Eberhardt, Ayan Paul, and Eibun Senaha for some technical help.  This research was supported in part by the Ministry of Science and Technology of Taiwan under Grant No.~MOST-108-2112-M-002-005-MY3.

\appendix

\section{List of Experimental References}
\label{appendix:exp list}

This appendix consists of several tables that list our experimental inputs.

\begin{table}[htb]
	\centering
		\begin{adjustbox}{width=0.9\columnwidth,center}
		\begin{tabular}{lr|c|c|c|c|c|c|c|}
			& & \textbf{$b\bar b$}  & \textbf{$WW$} & \textbf{$\tau\tau$} & \textbf{$ZZ$} & \textbf{$\gamma\gamma$} & \textbf{$Z\gamma$} & \textbf{$\mu\mu$} \\
			\hline
			\hline
			\multicolumn{2}{r|}{SM Br} & 57.5\% & 21.6\% & 6.3\% & 2.7\% & 2.3\textperthousand & 1.6\textperthousand & 0.2\textperthousand \\
			\hline
			\hline
			ggF$_8$ &87.2\%&--& \cellg\cite{ATLAS:2014aga,Chatrchyan:2013iaa}& \celly\cite{Aad:2015vsa,Chatrchyan:2014nva}& \cellg\cite{Aad:2014eva,Khachatryan:2014jba}& \cellg\cite{Aad:2014eha,Khachatryan:2014ira}&\cellr &\cellr\\
			ggF$_{13}$ &87.1\%&--& \cellg\cite{ATLAS-CONF-2020-027,Sirunyan:2018egh}& \celly\cite{ATLAS-CONF-2020-027,ATLAS-CONF-2018-021,Sirunyan:2017khh}& \cellg\cite{ATLAS-CONF-2020-027,Sirunyan:2017exp}& \cellg\cite{ATLAS-CONF-2020-027,Sirunyan:2018ouh,ATLAS-CONF-2018-028}& \cellr$8$ $\rm{TeV}$&\cellr$8$ $\rm{TeV}$\\
			\cline{1-7}
			VBF$_8$ &7.2\%&--& \cellg\cite{ATLAS:2014aga,Chatrchyan:2013iaa}& \celly\cite{Aad:2015vsa,Chatrchyan:2014nva}& \cellr\cite{Aad:2014eva,Khachatryan:2014jba}& \celly\cite{Aad:2014eha,Khachatryan:2014ira}&\cellr\cite{Aad:2015gba,Chatrchyan:2013vaa} & \cellr\cite{ATLAS:2016neq}\\
			VBF$_{13}$ &7.4\%& \cellr\cite{Aad:2020jym,CMS-PAS-HIG-16-003}& \cellg\cite{ATLAS-CONF-2020-045,Sirunyan:2018egh}& \celly\cite{ATLAS-CONF-2020-027,ATLAS-CONF-2018-021,Sirunyan:2017khh}& \cellg\cite{ATLAS-CONF-2020-027,Sirunyan:2017exp}& \cellg\cite{ATLAS-CONF-2020-027,Sirunyan:2018ouh,ATLAS-CONF-2018-028}&\cellr & \cellr\\
			\hline
			Vh$_8$ &5.1\%& \celly\cite{Aad:2014xzb,Chatrchyan:2013zna}& \cellr\cite{Aad:2015ona,Chatrchyan:2013iaa}& \cellr\cite{Aad:2015vsa,Chatrchyan:2014nva}& \cellr\cite{Aad:2014eva,Khachatryan:2014jba}& \celly\cite{Aad:2014eha,Khachatryan:2014ira}&\cellr & \cellr\\
			Vh$_{13}$ &4.4\%& \cellg\cite{Aad:2020jym,Sirunyan:2017elk}& \cellr\cite{ATLAS-CONF-2016-112,Sirunyan:2018egh}& \cellr\cite{Sirunyan:2017khh}& \celly\cite{ATLAS-CONF-2020-027,Sirunyan:2017exp}& \cellg\cite{ATLAS-CONF-2020-027,Sirunyan:2018ouh,ATLAS-CONF-2018-028}& \cellr$13$ $\rm{TeV}$&\cellr$13$ $\rm{TeV}$\\
			\cline{1-7}
			tth$_8$ &0.6\%& \cellr\cite{Aad:2015gra,Khachatryan:2014qaa}&--&--& \cellg\cite{Aad:2014eva,Khachatryan:2014jba}& \cellr\cite{Aad:2014eha,Khachatryan:2014ira}&\cellr\cite{ATLAS:2020qcv,Aaboud:2017uhw,CMS-PAS-HIG-19-014,Sirunyan:2018tbk} & \cellr\cite{ATLAS:2017eix,Sirunyan:2018hbu}\\
			tth$_{13}$ &1.0\%& \cellg\cite{Aad:2020jym,CMS-PAS-HIG-17-026,Sirunyan:2018ygk}& \cellr\cite{Aaboud:2017jvq,Sirunyan:2018egh,Sirunyan:2018shy}& \celly\cite{ATLAS-CONF-2020-027,Aaboud:2017jvq,Sirunyan:2018shy}& \cellr\cite{Aaboud:2017jvq,Sirunyan:2017exp,Sirunyan:2018shy}& \cellg\cite{ATLAS-CONF-2020-027,Sirunyan:2018ouh,ATLAS-CONF-2018-028}& \cellr&\cellr \\
			\hline
			\hline
			Vh$_2$ & &  \celly\cite{Aaltonen:2013ipa,Abazov:2013gmz} \\
			\hline
			tth$_2$ & &  \cellr\cite{Aaltonen:2013ipa} \\
			\hline
		\end{tabular}\end{adjustbox}\\[10pt]
		\scalebox{0.7}{\begin{tabular}{|c||c||c|}\hline \cellg $0<\hat\sigma<0.5$ & \celly$0.5\leq \hat\sigma \leq 1.0$ & \cellr$\hat\sigma>1.0$\\\hline\end{tabular} \quad ($\hat\sigma=\sigma_{\text{\tiny{min}}}/w$)}
		\caption{Higgs signal strength inputs applied in our fits. The Higgs decays are listed in separate columns, with the corresponding SM branching ratios shown in the second row. In lines three to ten, we cite the results from the LHC and Tevatron, ordered by production mechanism and $\sqrt{s}$. For the LHC data, we indicate the share of Higgs production in $pp$ collisions for each channel in the second column. The cell colors show the rough estimates on the current precision of the signal strength measurements according to the parameter $\hat{\sigma}$, which is defined as the ratio of the smallest uncertainty of all individual measurements in one table cell ($\sigma_{min}$) to the weight of the corresponding production channel ($w$). The green, yellow, and red cells denote the channels with $\hat{\sigma}$ less than 0.5, between 0.5 and 1, and greater than 1, respectively. For the $Z\gamma$ and $\mu\mu$ decays, no information of the individual production modes is available, and thus we assume the SM compositions in the second column to analyze them. In comparison with Ref.~\cite{Chiang:2018cgb}, the udpated 13-TeV analyses are quoted from Refs.~\cite{ATLAS-CONF-2020-027,ATLAS-CONF-2020-045,Aad:2020jym,CMS-PAS-HIG-17-026,CMS-PAS-HIG-19-014,ATLAS:2020qcv}.}
		\label{tab:signalstrengthinputs}
\end{table}

\begin{table}[htb]
	\centering
	\begin{adjustbox}{width=0.7\columnwidth,center}
		\begin{tabular}{|l|l|l|c|c|}
			\hline \textbf{Label} & \textbf{Channel} & \textbf{Experiment} & \textbf{Mass range {$[\mathrm{TeV}]$}} & \textbf{$\mathcal{L}$ {$\left[\mathrm{fb}^{-1}\right]$}}\\
			\hline \hline$A_{13 t}^{t t}$ & $t t \rightarrow \phi^{0} \rightarrow t t$ & ATLAS \hfill{\cite{ATLAS:2018alq}} & {$[0.4 , 1]$} & $36.1$ \\
			\hline$A_{13 b}^{t t}$ & $b b \rightarrow \phi^{0} \rightarrow t t$ & ATLAS \hfill{\cite{ATLAS-CONF-2016-104}} & {$[0.4 , 1]$} & $13.2$ \\
			\hline \hline$C_{8 b}^{b b}$ & $b b \rightarrow \phi^{0} \rightarrow b b$ & CMS \hfill{\cite{CMS:2015grx}} & {$[0.1 , 0.9]$} & $19.7$ \\
			\hline$C_{8}^{b b}$ & $g g \rightarrow \phi^{0} \rightarrow b b$ & CMS \hfill{\cite{CMS:2018kcg}} & {$[0.33 , 1.2]$} & $19.7$ \\
			\hline$C_{13}^{b b}$ & $p p \rightarrow \phi^{0} \rightarrow b b$ & CMS \hfill{\cite{CMS-PAS-HIG-16-025}} & {$[0.55 , 1.2]$} & $2.69$ \\
			\hline$C_{13 b}^{b b}$ & $b b \rightarrow \phi^{0} \rightarrow b b$ & CMS \hfill{\cite{CMS:2018hir}} & {$[0.3 , 1.3]$} & $35.7$ \\
			\hline\hline
			$A_{8}^{\tau \tau}$ & \multirow{2}{*}{$g g \rightarrow \phi^{0} \rightarrow \tau \tau$} & ATLAS \hfill{\cite{ATLAS:2014vhc}} & {$[0.09 , 1]$} & 20 \\
			$C_{8}^{\tau \tau}$ & & CMS \hfill{\cite{CMS-PAS-HIG-14-029}} & {$[0.09 , 1]$} & $19.7$ \\
			\hline$A_{8 b}^{\tau \tau}$ & \multirow{2}{*}{$b b \rightarrow \phi^{0} \rightarrow \tau \tau$} & ATLAS \hfill{\cite{ATLAS:2014vhc}} & {$[0.09 , 1]$} & 20 \\
			$C_{8 b}^{\tau \tau}$ & & CMS \hfill{\cite{CMS-PAS-HIG-14-029}} & {$[0.09 , 1]$} & $19.7$ \\
			\hline$A_{13}^{\tau \tau}$ & \multirow{2}{*}{$g g \rightarrow \phi^{0} \rightarrow \tau \tau$} & ATLAS \hfill{\cite{ATLAS:2017eiz}} & {$[0.2 , 2.25]$} & $36.1$ \\
			$C_{13}^{\tau \tau}$ & & CMS \hfill{\cite{CMS:2018rmh}} & {$[0.09 , 3.2]$} & $35.9$ \\
			\hline$A_{13 b}^{\tau \tau}$ & \multirow{2}{*}{$b b \rightarrow \phi^{0} \rightarrow \tau \tau$} & ATLAS \hfill{\cite{ATLAS:2017eiz}} & {$[0.2 , 2.25]$} & $36.1$ \\
			$C_{13 b}^{\tau \tau}$ & & CMS \hfill{\cite{CMS:2018rmh}} & {$[0.09 , 3.2]$} & $35.9$ \\
			\hline
	\end{tabular}\end{adjustbox}
	\caption{Neutral heavy Higgs boson searches relevant for the GM scalars with fermionic final states. $\phi^0 = H_1^0, H_3^0$.}
	\label{tab:neutral heavy higgs with fermions}
\end{table}

\begin{table}[htb]
	\centering
	\begin{adjustbox}{width=0.7\columnwidth,center}
		\begin{tabular}{|l|l|l|c|c|}
			\hline \textbf{Label} & \textbf{Channel} & \textbf{Experiment} & \textbf{Mass range {$[\mathrm{TeV}]$}} & \textbf{$\mathcal{L}$ {$\left[\mathrm{fb}^{-1}\right]$}}\\
			\hline$A_{8}^{\gamma \gamma}$ & $g g \rightarrow \phi^{0} \rightarrow \gamma \gamma$ & ATLAS \hfill{\cite{ATLAS:2014jdv}} & {$[0.065 , 0.6]$} & $20.3$ \\
			\hline$A_{13}^{\gamma \gamma}$ & $p p \rightarrow \phi^{0} \rightarrow \gamma \gamma$ & ATLAS \hfill{\cite{ATLAS:2017ayi}} & {$[0.2 , 2.7]$} & $36.7$ \\
			\hline$C_{13}^{\gamma \gamma}$ & $g g \rightarrow \phi^{0} \rightarrow \gamma \gamma$ & CMS \hfill{\cite{CMS:2016kgr}} & {$[0.5 , 4]$} & $35.9$ \\
			\hline \hline
			$A_{8}^{Z \gamma}$ & \multirow{2}{*}{$p p \rightarrow \phi^0 \rightarrow Z \gamma \rightarrow (ll) \gamma$}& ATLAS \hfill{\cite{ATLAS:2014lfk}} & {$[0.2 , 1.6]$} & 20.3 \\
			$C_{8}^{Z \gamma}$ & & CMS \hfill{\cite{CMS-PAS-HIG-16-014}} & {$[0.2 , 1.2]$} & $19.7$ \\
			\hline$A_{13}^{\ell \gamma}$ & $g g \rightarrow \phi^{0} \rightarrow Z \gamma[\rightarrow(\ell \ell) \gamma]$ & ATLAS \hfill{\cite{ATLAS:2017zdf}} & {$[0.25 , 2.4]$} & $36.1$ \\
			\hline$A_{13}^{q q \gamma}$ & $g g \rightarrow \phi^{0} \rightarrow Z \gamma[\rightarrow(q q) \gamma]$ & ATLAS \hfill{\cite{ATLAS:2018sxj}} & {$[1 , 6.8]$} & $36.1$ \\
			\hline$C_{8+13}^{Z \gamma}$ & $g g \rightarrow \phi^{0} \rightarrow Z \gamma$ & CMS \hfill{\cite{CMS:2017dyb}} & {$[0.35 , 4]$} & $35.9$ \\
			\hline \hline
			$A_{8}^{Z Z}$ & $g g \rightarrow \phi^{0} \rightarrow Z Z$ & ATLAS \hfill{\cite{ATLAS:2015pre}} & {$[0.14 , 1]$} & $20.3$ \\
			\hline$A_{8 V}^{Z Z}$ & $V V \rightarrow \phi^{0} \rightarrow Z Z$ & ATLAS \hfill{\cite{ATLAS:2015pre}} & {$[0.14 , 1]$} & $20.3$ \\
			\hline$A_{13,1}^{2\ell2 L}$ & $g g \rightarrow \phi^{0} \rightarrow Z Z[\rightarrow(\ell \ell)(\ell \ell, \nu \nu)]$ & ATLAS \hfill{\cite{ATLAS:2017tlw}} & {$[0.2 , 1.2]$} & $36.1$ \\
			\hline$A_{13 V,1}^{2 \ell 2 L}$ & $V V \rightarrow \phi^{0} \rightarrow Z Z[\rightarrow(\ell \ell)(\ell \ell, \nu \nu)]$ & ATLAS \hfill{\cite{ATLAS:2017tlw}} & {$[0.2 , 1.2]$} & $36.1$ \\
			\hline$A_{13}^{2 L 2 q}$ & $g g \rightarrow \phi^{0} \rightarrow Z Z[\rightarrow(\ell \ell, \nu \nu)(q q)]$ & ATLAS \hfill{\cite{ATLAS:2017otj}} & {$[0.3 , 3]$} & $36.1$ \\
			\hline$A_{13 V}^{2 L 2 q}$ & $V V \rightarrow \phi^{0} \rightarrow Z Z[\rightarrow(\ell \ell, \nu \nu)(q q)]$ & ATLAS \hfill{\cite{ATLAS:2017otj}} & {$[0.3 , 3]$} & $36.1$ \\
			\hline$C_{13}^{2\ell2 X}$ & $p p \rightarrow \phi^{0} \rightarrow Z Z[\rightarrow(\ell \ell)(q q, \nu \nu, \ell \ell)]$ & CMS \hfill{\cite{CMS:2018amk}} & {$[0.13 , 3]$} & $35.9$ \\
			\hline 
			${C}_{13}^{2 q 2 \nu}$ & $p p \rightarrow \phi^{0} \rightarrow Z Z[\rightarrow(q q)(\nu \nu)]$ & CMS \hfill{\cite{CMS:2018ygj}} & {$[1 , 4]$} & $35.9$ \\
			\hline
			\hline$A_{8}^{W W}$ & $g g \rightarrow \phi^{0} \rightarrow W W$ & ATLAS \hfill{\cite{ATLAS:2015iie}} & {$[0.3 , 1.5]$} & $20.3$ \\
			\hline$A_{8 V}^{W W}$ & $V V \rightarrow \phi^{0} \rightarrow W W$ & ATLAS \hfill{\cite{ATLAS:2015iie}} & {$[0.3 , 1.5]$} & $20.3$ \\
			\hline$A_{13}^{2(\ell \nu)}$ & $g g \rightarrow \phi^{0} \rightarrow W W[\rightarrow(e \nu)(\mu \nu)]$ & ATLAS \hfill{\cite{ATLAS:2017uhp}} & {$[0.25 , 4]$} & $36.1$ \\
			\hline$A_{13 V}^{2(\ell \nu)}$ & $V V \rightarrow \phi^{0} \rightarrow W W[\rightarrow(e \nu)(\mu \nu)]$ & ATLAS \hfill{\cite{ATLAS:2017uhp}} & {$[0.25 , 3]$} & $36.1$ \\
			\hline$C_{13}^{2(\ell \nu)}$ & $(g g+V V) \rightarrow \phi^{0} \rightarrow W W \rightarrow(\ell \nu)(\ell \nu)$ & CMS \hfill{\cite{CMS-PAS-HIG-16-023}} & {$[0.2 , 1]$} & $2.3$ \\
			\hline$A_{13}^{\ell \nu 2 q}$ & $g g \rightarrow \phi^{0} \rightarrow W W[\rightarrow(\ell \nu)(q q)]$ & ATLAS \hfill{\cite{ATLAS:2017jag}} & {$[0.3 , 3]$} & $36.1$ \\
			\hline$A_{13 V}^{\ell \nu 2 q}$ & $V V \rightarrow \phi^{0} \rightarrow W W[\rightarrow(\ell \nu)(q q)]$ & ATLAS \hfill{\cite{ATLAS:2017jag}} & {$[0.3 , 3]$} & $36.1$ \\
			\hline \hline$C_{8}^{V V}$ & $p p \rightarrow \phi^{0} \rightarrow V V$ & CMS \hfill{\cite{CMS:2015hra}} & {$[0.145 , 1]$} & $24.8$ \\
			\hline
	\end{tabular}\end{adjustbox}
	\caption{Neutral heavy Higgs boson searches relevant for the GM scalars with vector boson final
		states. $\phi^0 = H_1^0, H_3^0, H_5^0$ and $\ell = e, \mu$.}
	\label{tab:neutral heavy higgs with bosons}
\end{table}

\begin{table}[htb]
	\centering
	\begin{adjustbox}{width=0.7\columnwidth,center}
		\begin{tabular}{|l|l|l|c|c|}
			\hline \textbf{Label} & \textbf{Channel} & \textbf{Experiment} & \textbf{Mass range {$[\mathrm{TeV}]$}} & \textbf{$\mathcal{L}$ {$\left[\mathrm{fb}^{-1}\right]$}}\\
			\hline \hline$A_{8}^{h h}$ & $g g \rightarrow H_{1}^{0} \rightarrow h h$ & ATLAS \hfill{\cite{ATLAS:2015sxd}} & {$[0.26 , 1]$} & $20.3$ \\
			\hline$C_{8}^{4 b}$ & $p p \rightarrow H_{1}^{0} \rightarrow h h \rightarrow(b b)(b b)$ & CMS \hfill{\cite{CMS:2015jal}} & {$[0.27 , 1.1]$} & $17.9$ \\
			\hline$C_{8}^{2 \gamma 2 b}$ & $p p \rightarrow H_{1}^{0} \rightarrow h h \rightarrow(b b)(\gamma \gamma)$ & CMS \hfill{\cite{CMS:2016cma}} & {$[0.260 , 1.1]$} & $19.7$ \\
			\hline$C_{8 g}^{2 b 2 \tau}$ & $g g \rightarrow H_{1}^{0} \rightarrow h h \rightarrow(b b)(\tau \tau)$ & CMS \hfill{\cite{CMS:2015uzk}} & {$[0.26 , 0.35]$} & $19.7$ \\
			\hline$C_{8}^{2 b 2 \tau}$ & $p p \rightarrow H_{1}^{0} \rightarrow h h[\rightarrow(b b)(\tau \tau)]$ & CMS \hfill{\cite{CMS:2017yfv}} & {$[0.35 , 1]$} & $18.3$ \\
			\hline$A_{13}^{4 b}$ & \multirow{2}{*}{$p p \rightarrow H_{1}^{0} \rightarrow h h \rightarrow(b b)(b b)$} & ATLAS \hfill{\cite{ATLAS:2018rnh}} & {$[0.26 , 3]$} & $36.1$ \\
			$C_{13}^{4 b}$ & & CMS \hfill{\cite{CMS:2018qmt}} & {$[0.26 , 1.2]$} & $35.9$ \\
			\hline$A_{13}^{2\gamma2b}$ & $p p \rightarrow H_{1}^{0} \rightarrow h h[\rightarrow(b b)(\gamma \gamma)]$ & ATLAS \hfill{\cite{ATLAS:2018dpp}} & {$[0.26 , 1]$} & $36.1$ \\
			$C_{13}^{2 \gamma 2 b}$ & $p p \rightarrow H_{1}^{0} \rightarrow h h \rightarrow(b b)(\gamma \gamma)$ & CMS \hfill{\cite{CMS:2018tla}} & {$[0.25 , 0.9]$} & $35.9$ \\
			\hline$A_{13}^{2b2\tau}$ & \multirow{2}{*}{$p p \rightarrow H_{1}^{0} \rightarrow h h \rightarrow(b b)(\tau \tau)$} & ATLAS \hfill{\cite{ATLAS:2018uni}} & {$[0.26 , 1]$} & $36.1$ \\
			$C_{13,1}^{2 b 2 \tau}$ & & CMS \hfill{\cite{CMS:2017hea}}& {$[0.25 , 0.9]$} & $35.9$ \\
			$C_{13,2}^{2 b 2 \tau}$ & $p p \rightarrow H_{1}^{0} \rightarrow h h[\rightarrow(b b)(\tau \tau)]$ & CMS \hfill{\cite{CMS:2018kaz}} & {$[0.9 , 4]$} & $35.9$ \\
			\hline$C_{13}^{2b2 V}$ & $p p \rightarrow H_{1}^{0} \rightarrow h h \rightarrow(b b)(V V \rightarrow \ell \nu \ell \nu)$ & CMS \hfill{\cite{CMS:2017rpp}} & {$[0.26 , 0.9]$} & $35.9$ \\
			\hline$A_{13}^{2 \gamma 2 W}$ & $g g \rightarrow H_{1}^{0} \rightarrow h h \rightarrow(\gamma \gamma)(W W)$ & ATLAS \hfill{\cite{ATLAS:2018hqk}} & {$[0.26 , 0.5]$} & $36.1$ \\
			\hline
			\hline$A_{8}^{b b Z}$ & $g g \rightarrow H_{3}^{0} \rightarrow h Z \rightarrow(b b) Z$ & ATLAS \hfill{\cite{ATLAS:2015kpj}} & {$[0.22 , 1]$} & $20.3$ \\
			\hline$C_{8}^{2 b 2 \ell}$ & $g g \rightarrow H_{3}^{0} \rightarrow h Z \rightarrow(b b)(\ell \ell)$ & CMS \hfill{\cite{CMS:2015flt}} & {$[0.225 , 0.6]$} & $19.7$ \\
			\hline$A_{8}^{\tau \tau Z}$ & $g g \rightarrow H_{3}^{0} \rightarrow h Z \rightarrow(\tau \tau) Z$ & ATLAS \hfill{\cite{ATLAS:2015kpj}} & {$[0.22 , 1]$} & $20.3$ \\
			\hline$C_{8}^{2 \tau 2 \ell}$ & $g g \rightarrow H_{3}^{0} \rightarrow h Z \rightarrow(\tau \tau)(\ell \ell)$ & CMS \hfill{\cite{CMS:2015uzk}} & {$[0.22 , 0.35]$} & $19.7$ \\
			\hline$A_{13}^{b b Z}$ & & ATLAS \hfill{\cite{ATLAS:2017xel}} & {$[0.2 , 2]$} & $36.1$ \\
			$C_{13,1}^{b b Z}$ & $g g \rightarrow H_{3}^{0} \rightarrow h Z \rightarrow(b b) Z$ & CMS \hfill{\cite{CMS-PAS-HIG-18-005}} & {$[0.22 , 0.8]$} & $35.9$ \\
			$C_{13,2}^{b b Z}$ & & CMS \hfill{\cite{CMS:2018ljc}} & {$[0.8 , 2]$} & $35.9$ \\
			\hline$A_{13 b}^{b b Z}$ & & ATLAS \hfill{\cite{ATLAS:2017xel}} & {$[0.2 , 2]$} & $36.1$ \\
			$C_{13 b, 1}^{b b Z}$ & $b b \rightarrow H_{3}^{0} \rightarrow h Z \rightarrow(b b) Z$ & CMS \hfill{\cite{CMS-PAS-HIG-18-005}} & {$[0.22 , 0.8]$} & $35.9$ \\
			$C_{13 b, 2}^{b b Z}$ & & CMS \hfill{\cite{CMS:2018ljc}} & {$[0.8 , 2]$} & $35.9$ \\
			\hline \hline$C_{8}^{\phi Z}$ & $p p \rightarrow \phi^{0} \rightarrow \phi^{0 \prime} Z \rightarrow(b b)(\ell \ell)$ & CMS \hfill{\cite{CMS:2016xnc}} & {$[0.13 , 1]$} & $19.8$ \\
			\hline$A_{13}^{\phi Z}$ & $g g \rightarrow H_{3}^{0} \rightarrow H_{1}^{0} Z \rightarrow(b b) Z$ & ATLAS \hfill{\cite{ATLAS:2018oht}} & {$[0.13 , 0.8]$} & $36.1$ \\
			\hline \hline$A_{13 b}^{\phi Z}$ & $b b \rightarrow H_{3}^{0} \rightarrow H_{1}^{0} Z \rightarrow(b b) Z$ & ATLAS \hfill{\cite{ATLAS:2018oht}} & {$[0.13 , 0.8]$} & $36.1$ \\
			\hline
	\end{tabular}\end{adjustbox}
	\caption{Neutral heavy Higgs boson searches at the LHC relevant for the GM scalars with final
		states including Higgs bosons. $\phi^0 = H_1^0 , H_3^0 , H_5^0$ , $\phi^{0\prime} = H_1^0 , H_3^0$ , $V = W, Z$ and $\ell = e, \mu$.}
	\label{tab:neutral heavy higgs with higgs}
\end{table}

\begin{table}[htb]
	\centering
	\begin{adjustbox}{width=0.7\columnwidth,center}
		\begin{tabular}{|l|l|l|c|c|}
			\hline \textbf{Label} & \textbf{Channel} & \textbf{Experiment} & \textbf{Mass range {$[\mathrm{TeV}]$}} & \textbf{$\mathcal{L}$ {$\left[\mathrm{fb}^{-1}\right]$}}\\
			\hline \hline$A_{8}^{\tau \nu}$ & $p p \rightarrow H_{3}^{\pm} \rightarrow \tau^{\pm} \nu$ & ATLAS \hfill{\cite{ATLAS:2014otc}} & {$[0.18 , 1]$} & $19.5$ \\
			\hline$C_{8}^{\tau \nu}$ & $p p \rightarrow H_{3}^{+} \rightarrow \tau^{+} \nu$ & CMS \hfill{\cite{CMS:2015lsf}} & {$[0.18 , 0.6]$} & $19.7$ \\
			\hline$A_{13}^{\tau V}$ & \multirow{2}{*}{$p p \rightarrow H_{3}^{\pm} \rightarrow \tau^{\pm} \nu$} & ATLAS \hfill{\cite{ATLAS:2018gfm}} & {$[0.15 , 2]$} & $36.1$ \\
			$C_{13}^{\tau \nu}$ & & CMS \hfill{\cite{CMS-PAS-HIG-16-031}} & {$[0.18 , 3]$} & $12.9$ \\
			\hline \hline$A_{8}^{t b}$ & $p p \rightarrow H_{3}^{\pm} \rightarrow t b$ & ATLAS \hfill{\cite{ATLAS:2015nkq}} & {$[0.2 , 0.6]$} & $20.3$ \\
			\hline$C_{8}^{t b}$ & $p p \rightarrow H_{3}^{+} \rightarrow t \bar{b}$ & CMS \hfill{\cite{CMS:2015lsf}} & {$[0.18 , 0.6]$} & $19.7$ \\
			\hline$A_{13}^{t b}$ & $p p \rightarrow H_{3}^{\pm} \rightarrow t b$ & ATLAS \hfill{\cite{ATLAS:2018ntn}} & {$[0.2 , 2]$} & $36.1$ \\
			\hline \hline$A_{8}^{W Z}$ & $W Z \rightarrow H_{5}^{\pm} \rightarrow W Z[\rightarrow(q q)(\ell \ell)]$ & ATLAS \hfill{\cite{ATLAS:2015edr}} & {$[0.2 , 1]$} & $20.3$ \\
			\hline$A_{13}^{W Z}$ & & ATLAS \hfill{\cite{ATLAS:2018iui}} & {$[0.2 , 0.9]$} & $36.1$ \\
			$C_{13,1}^{W Z}$ & $W Z \rightarrow H_{5}^{\pm} \rightarrow W Z[\rightarrow(\ell \nu)(\ell \ell)]$ & CMS \hfill{\cite{CMS:2017fgp}} & {$[0.2 , 0.3]$} & $15.2$ \\
			$C_{13,2}^{W Z}$ & & CMS \hfill{\cite{CMS-PAS-SMP-18-001}} & {$[0.3 , 2]$} & $35.9$ \\
			\hline \hline$A_{13,1}^{4 W}$ & $p p \rightarrow H_{5}^{\pm \pm} H_{5}^{\mp \mp} \rightarrow\left(W^{\pm} W^{\pm}\right)\left(W^{\mp} W^{\mp}\right)$ & ATLAS \hfill{\cite{ATLAS:2018ceg}} & {$[0.2 , 0.7]$} & $36.1$ \\
			\hline \hline$C_{8}^{\ell^\pm \ell^{\pm}}$ & $V V \rightarrow H_{5}^{\pm \pm} \rightarrow W^{\pm} W^{\pm}\left[\rightarrow\left(\ell^{\pm} \nu\right)\left(\ell^{\pm} \nu\right)\right]$ & CMS \hfill{\cite{CMS:2014mra}} & {$[0.2 , 0.8]$} & $19.4$ \\
			\hline$C_{13,1}^{\ell^{\pm} \ell^{\pm}}$ & $V V \rightarrow H_{5}^{\pm \pm} \rightarrow W^{\pm} W^{\pm}\left[\rightarrow\left(\ell^{\pm} \nu\right)\left(\ell^{\pm} \nu\right)\right]$ & CMS \hfill{\cite{CMS:2017fhs}} & {$[0.2 , 1.0]$} & $35.9$ \\
			\hline
	\end{tabular}\end{adjustbox}
	\caption{Charged heavy Higgs boson searches at the LHC relevant for the singly and doubly charged scalars in the GM model, with $V = W, Z$ and $\ell = e, \mu$.}
	\label{tab:charged heavy higgs with charged scalars}
\end{table}

\begin{table}[htb]
	\centering
	\begin{adjustbox}{width=0.7\columnwidth,center}
		\begin{tabular}{|l|l|l|c|c|}
			\hline \textbf{Label} & \textbf{Channel} & \textbf{Experiment} & \textbf{Mass range {$[\mathrm{TeV}]$}} & \textbf{$\mathcal{L}$ {$\left[\mathrm{fb}^{-1}\right]$}}\\
			\hline
			\hline
			$C^{h}_{13}$ & $p p \rightarrow \phi^{0} \rightarrow h h$ & CMS \cite{CMS:2018ipl} & [0.27, 3] & 35.9\\
			\hline
			$C^{\mu}_{13b}$ & $b b \rightarrow \phi^{0} \rightarrow \mu \mu$ & ATLAS \cite{ATLAS:2019odt} & [0.2,1] & 36.1\\
			$C^{\mu}_{13}$ & $g g \rightarrow \phi^{0} \rightarrow \mu \mu$ & ATLAS \cite{ATLAS:2019odt} & [0.2,1] & 36.1\\
			\hline
			$A^{b}_{13b}$ & $b b \rightarrow \phi^{0} \rightarrow b b$ & ATLAS \cite{ATLAS:2019tpq}& [0.45,1.4] & 27.8\\
			\hline
			$C^{\mu}_{13b}$ & $b b \rightarrow \phi^{0} \rightarrow \mu \mu$ & CMS \cite{CMS:2019mij}& [0.13,1] & 36.1\\
			$C^{\mu}_{13}$ & $g g \rightarrow \phi^{0} \rightarrow \mu \mu$ & CMS \cite{CMS:2019mij}& [0.13,1] & 36.1\\
			\hline
			$C^{l \tau}_{13}$ & $g g \rightarrow H_{3}^{0} \rightarrow \ell^{\pm} \ell^\mp \tau^{\pm} \tau^{\mp}$ & CMS \cite{CMS:2019kca} & [0.22, 0.4] & 35.9\\
			\hline
			$C^{W}_{13}$ & $p p \rightarrow \phi^{0} \rightarrow W^{\pm} W^{\mp}$ & CMS \cite{CMS:2019bnu} & [0.2, 3] & 35.9\\
			$C^{W}_{13V}$ & $V V \rightarrow \phi^{0} \rightarrow W^{\pm} W^{\mp}$ & CMS \cite{CMS:2019bnu} & [0.2, 3] & 35.9\\
			\hline
			$A^{\tau}_{13}$ & $g g \rightarrow \phi^0 \rightarrow \tau^{\pm} \tau^{\mp}$ & ATLAS \cite{ATLAS:2020zms} & [0.2, 2.5] & 139\\
			$A^{\tau}_{13b}$ & $b b \rightarrow \phi^0 \rightarrow \tau^{\pm} \tau^{\mp}$ & ATLAS \cite{ATLAS:2020zms} & [0.2, 2.5] & 139\\
			\hline
			$A^{2\ell 2L}_{13,2}$ & $g g \rightarrow \phi^0 \rightarrow Z Z \rightarrow (\ell\ell\ell\ell)+(\ell\ell\nu\nu)$ & ATLAS \cite{ATLAS:2020tlo} & [0.21, 2] & 139\\
			$A^{2\ell 2L}_{13V,2}$ & $V V \rightarrow \phi^0 \rightarrow Z Z \rightarrow (\ell\ell\ell\ell)+(\ell\ell\nu\nu)$ & ATLAS \cite{ATLAS:2020tlo} & [0.21, 2] & 139\\
			\hline
	\end{tabular}
	\end{adjustbox}
	\caption{Newly added direct searches for neutral heavy Higgs bosons.}
	\label{tab:new neutral direct searches}
\end{table}
\begin{table}[htb]
	\centering
	\begin{adjustbox}{width=0.7\columnwidth,center}
		\begin{tabular}{|l|l|l|c|c|}
			\hline \textbf{Label} & \textbf{Channel} & \textbf{Experiment} & \textbf{Mass range {$[\mathrm{TeV}]$}} & \textbf{$\mathcal{L}$ {$\left[\mathrm{fb}^{-1}\right]$}}\\
			\hline
			\hline
			$C^{\tau\nu}_{13}$ & $p p \rightarrow H_{3}^{\pm} \rightarrow \tau^{\pm} \nu$ & CMS \cite{CMS:2019bfg} & [0.08, 3] & 35.9\\
			\hline
			$C^{tb}_{13}$ & $p p \rightarrow H_{3}^{\pm} \rightarrow t b$ & CMS \cite{CMS:2020imj} & [0.2,3] & 35.9\\
			\hline
			$A^{4W}_{13,2}$ & $p p \rightarrow H_5^{\pm\pm}H_5^{\mp\mp} \rightarrow \left(W^\pm W^\pm\right) \left(W^\mp W^\mp\right)$ & ATLAS \cite{ATLAS:2021jol}& [0.2,0.6] & 139\\
			$A^{3W1Z}_{13}$ & $p p \rightarrow H_5^{\pm\pm}H_{3,5}^{\mp} \rightarrow W^\pm W^\pm W^\mp Z$ & ATLAS \cite{ATLAS:2021jol}& [0.2,0.6] & 139\\
			\hline
			$C^{\ell^\pm \ell^\pm}_{13,2}$ & $V V \rightarrow H_{5}^{\pm\pm} \rightarrow W^{\pm} W^\pm\left[\rightarrow\left(\ell \nu\right)\left(\ell \nu\right)\right]$ & CMS \cite{CMS:2021wlt} & [0.2, 3] & 137\\
			$C^{WZ}_{13,3}$ & $W^{\pm} Z \rightarrow H_{5}^{\pm} \rightarrow W^{\pm} Z[\rightarrow(\ell \nu)(\ell \ell)]$ & CMS \cite{CMS:2021wlt} & [0.2, 3] & 137\\
			\hline
			
	\end{tabular}\end{adjustbox}
	\caption{Newly added direct searches for heavy charged Higgs bosons}
	\label{tab:new charged direct searches}
\end{table}


\clearpage
%


\end{document}